\documentstyle[preprint,aps,epsf,tighten]{revtex}	

\begin{document}
%
%
\preprint{$
\begin{array}{l}
\mbox{BA-00-17}\\
\mbox{FERMILAB-Pub-00/068-T}\\
\mbox{March 2000}\\[0.5in]
\end{array}
$}
\title{Construction of a Minimal Higgs $SO(10)$ SUSY GUT Model}
\author{Carl H. Albright}
\address{Department of Physics, Northern Illinois University, DeKalb, IL
60115\\
       and\\
Fermi National Accelerator Laboratory, P.O. Box 500, Batavia, IL
60510\footnote{electronic address: albright@fnal.gov}}
\author{S.M. Barr}
\address{Bartol Research Institute,
University of Delaware, Newark, DE 19716\footnote{electronic address:
smbarr@bartol.udel.edu}}
\maketitle
%
\begin{abstract}
%
A full account is given of the procedure used by the authors to construct 
an $SO(10)$ supersymmetric grand unified model of the fermion 
mass matrices.  Various features of the model which gives remarkably 
accurate results for the quark and lepton masses and mixings were presented
earlier in separate publications.  The construction of the matrices is 
first discussed in the framework of effective operators, from which one 
naturally obtains the maximal $\nu_\mu - \nu_\tau$ mixing, while the 
small angle or maximal mixing solutions for the solar neutrinos depend
upon the nature of the Majorana matrix.  A set of Higgs and fermion superfields 
is then introduced from which the Higgs and Yukawa superpotentials uniquely
give the structure of the mass matrices previously obtained.  The right-handed
Majorana matrix arises from one Higgs field coupling to several pairs
of superheavy conjugate neutrino singlets.  For the simple version considered,
10 input parameters accurately yield the 20 masses and mixings of the  
quarks and leptons, and the 3 masses of the right-handed neutrinos.
\end{abstract}
%
\newpage
%
%

\section{INTRODUCTION}

In several recent papers [1-4] we have developed a highly 
predictive model of quark and lepton masses based on the grand unified 
group $SO(10)$. This model grew out of our attempt \cite{ab1} to construct a 
realistic grand unified theory (GUT) in which $SO(10)$ was broken down 
to the standard model in the simplest possible, or ``minimal" way 
\cite{b-r}.  In this model there emerged a new mechanism based on certain
well-known features of $SU(5)$ for explaining 
the large mixing between the mu and tau neutrinos that is seen at 
SuperKamiokande\cite{atm}.  In \cite{ab1,abb} we gave the structure
of the quark and lepton mass matrices for the second and third families,
treating the first family as massless. In \cite{ab4}, it was shown how
to extend the model to include the first family, which leads to several 
interesting predictions. In \cite{ab5}, it was observed that 
the mixing of the electron neutrino very naturally falls
either within the range $0.004 \leq \sin^2 2 \theta_{e \mu} 
\leq 0.008$, corresponding to the small angle MSW solution \cite{MSW}
of the solar neutrino problem, or very near to the value $\sin^2 
\theta_{e \mu} = 1$, corresponding to what is called ``bimaximal mixing".

In this paper we present the model in fuller detail, especially in regard
to neutrino phenomenology, and to the structure of the Higgs 
sector, Yukawa interactions, and flavor symmetries that underlie the 
quark and lepton mass matrices.

The paper is organized as follows. In Section II we discuss in general 
terms, that is apart from a particular model, our mechanism for 
explaining the large mixing of $\nu_{\mu}$ and $\nu_{\tau}$. In Section 
III, we explain what we mean by a ``minimal" $SO(10)$ breaking scheme, 
and show how such minimal breaking and the requirements of simplicity 
lead one naturally to a certain form for the mass terms of the heavier 
two families of quarks and leptons. We then observe that this form 
realizes the general mechanism for large $\nu_{\mu}-\nu_{\tau}$ mixing 
described in the previous Section. It is important to emphasize that 
this mechanism emerged not from an attempt to explain neutrino 
phenomenology, but from other considerations entirely,
in particular the attempt to simplify the Higgs structure of $SO(10)$.
It is most interesting that the same mechanism has also independently  
been found by other groups attempting to make sense of neutrino 
phenomenology. In Section IV, it will be explained how this model is 
best extended to the first family of quarks and leptons, and how this 
gives rise to several distinctive predictions. Accurate analytic 
expressions for the predictions at the GUT scale will be presented. In 
Section V, the neutrino sector will be examined in detail. It will be 
seen how either the small-angle MSW solution of the solar neutrino 
problem or bimaximal mixing can result with equal simplicity. Finally, 
in Section VI, a concrete model, including all the details of flavor 
symmetries and of the Higgs and Yukawa superpotentials, will be 
presented, showing that the basic scheme is technically natural.

\section{MECHANISM FOR LARGE $\nu_\mu - \nu_\tau$ MIXING}

Before explaining our mechanism, it will be helpful to explain why
the observed large mixing of $\nu_\mu$, presumably with 
$\nu_{\tau}$, has been a theoretical puzzle. The basic reason is simple:
the mixing that is seen between the quarks of the second and third 
families is described by a
small mixing angle, namely $V_{cb} \cong 0.04$, and therefore it was
expected that the mixing between the second and third family of leptons 
would also be small. 

The grounds for this expectation were twofold. First, there is the
empirical fact that the masses of the quarks and leptons exhibit roughly
similar ``hierarchical" patterns, and therefore it was natural to assume 
that their mixing angles would be similar also.
Second, the most promising theoretical 
approaches to understanding the pattern of quark and lepton masses,
namely grand unification and flavor symmetry, tend to treat quarks and
leptons in similar ways. For instance, small quark mixing angles might
suggest an underlying fundamental ``family symmetry" that is weakly 
broken, in which case the lepton mixings might be expected also to be 
small. And in grand unification based on $SO(10)$ there is a close 
connection between the quark and lepton mass matrices. 

There are actually two puzzles associated with the mixing of the second 
and third families: First, why is the lepton mixing $|U_{\mu 3}| \sim 0.7$ 
so large? And, second, why is the quark 
mixing $V_{cb} \cong 0.04$ so small? What we mean by saying that these 
are distinct puzzles is that they are both unexpected within the most 
commonly assumed framework for explaining quark and lepton masses, the 
Weinberg-Wilczek-Zee-Fritzsch (WWZF) idea \cite{w-w-z-f}.

The WWZF idea was that 
the Cabibbo angle could be understood if the mass matrices of the first 
and second families of quarks had the following form:

\begin{equation}
L_{mass} = (\overline{u_R}, \overline{c_R}) \left( 
\begin{array}{cc} 0 & b \\ b & a \end{array} \right) \left(
\begin{array}{c} u_L \\ c_L \end{array} \right)
+ (\overline{d_R}, \overline{s_R}) \left( 
\begin{array}{cc} 0 & b' \\ b' & a' \end{array} \right) \left(
\begin{array}{c} d_L \\ s_L \end{array} \right).
\end{equation}

\noindent
This gives $\left| m_d/m_s \right| \cong \left| b'/a' \right|^2$, 
$\left| m_u/m_c \right| \cong \left| b/a \right|^2$, and
$V_{us} \cong b'/a' - b/a$, and thus the famous relation

\begin{equation}
V_{us} \cong \sqrt{m_d/m_s} - e^{i \alpha} \sqrt{m_u/m_c}.
\end{equation}

\noindent
Since $\left| V_{us} \right| \cong 0.22$, $\sqrt{m_d/m_s} \cong 0.22$, 
and $\sqrt{m_u/m_c} \cong 0.07$, this relation is satisfied for
$\alpha \sim \pm \pi/2$. If we apply the same idea to the leptons of the
first two families we get

\begin{equation}
U_{e 2} \cong \sqrt{m_e/m_{\mu}} - e^{i \beta} \sqrt{m_{\nu_1}/
m_{\nu_2}}.
\end{equation}

\noindent
The second term on the right is not known, but if it is assumed to be
small one has the rough prediction that $U_{e 2} \sim \sqrt{m_e/m_{\mu}}
\cong 0.07$. This could be consistent with the small angle MSW
solution of the solar neutrino problem, which requires that
$U_{e 2} \sim 0.04$. Thus the WWZF idea appears to 
work well where it was originally applied, namely to the first and 
second families.

Fritzsch \cite{Fritzsch} later extended this idea to explain the mixing 
of the third family. If a WWZF form is assumed to hold for the second and 
third family, i.e., if one takes $(u, c) \longrightarrow (c, t)$ and
$(d, s) \longrightarrow (s, b)$ in Eq. (1), one obtains 

\begin{equation}
V_{cb} \cong \sqrt{m_s/m_b} - e^{i \gamma} \sqrt{m_c/m_t}
\end{equation}

\noindent
and

\begin{equation}
U_{\mu 3} \cong \sqrt{m_{\mu}/m_{\tau}} - e^{i \delta} 
\sqrt{m_{\nu_2}/m_{\nu_3}}.
\end{equation} 

\noindent
Since $\sqrt{m_s/m_b} \cong 0.14$, and $\sqrt{m_c/m_t} \cong 0.04$,
one sees that the observed value of $V_{cb} \cong 0.04$ is too small
by a factor of three or so. Assuming that the neutrino mass ratio
in Eq. (5) is small, and given that $\sqrt{m_{\mu}/m_{\tau}} \cong 0.24$,
one sees that the nearly maximal value of $U_{\mu 3} \sim 1/\sqrt{2}
\cong 0.7$ that is observed is too large by a factor of three or so.

These Eqs. (2-5) are based on the assumption of a hierarchical and 
symmetric form for the mass matrices.  A key feature in our mechanism 
for understanding the large mixing of the tau neutrino is that it 
involves highly asymmetric mass matrices. As we shall see, the 
assumption of asymmetric mass matrices naturally explains why 
$U_{\mu 3}$ is larger than the Fritzsch value and $V_{cb}$ is smaller 
than the Fritzsch value by approximately the same factor. 

Consider, a toy model with $SU(5)$ symmetry, which has a set of Yukawa 
terms of the following form: $\lambda_{33}(\overline{{\bf 5}}_3 
{\bf 10}_3)
\overline{{\bf 5}}_H + \lambda_{23}(\overline{{\bf 5}}_2 {\bf 10}_3) 
\overline{{\bf 5}}_H + \lambda_{32}(\overline{{\bf 5}}_3 {\bf 10}_2) 
\overline{{\bf 5}}_H$, with $\lambda_{32} \ll \lambda_{23} \sim 
\lambda_{33}$ and the subscript $H$ denoting a Higgs representation.
These terms yield the following mass matrices for
the second and third families of down quarks and charged leptons:

\begin{equation}
(\overline{d_{2R}}, \overline{d_{3R}}) \left( \begin{array}{cc}
0 & \sigma \\ \epsilon & 1 \end{array} \right)  \left( 
\begin{array}{c} d_{2L} \\ d_{3L} \end{array} \right) M_D +
(\overline{l_{2R}}, \overline{l_{3R}}) \left( \begin{array}{cc}
0 & \epsilon \\ \sigma & 1 \end{array} \right) \left( 
\begin{array}{c} l_{2L} \\ l_{3L} \end{array} \right) M_D,
\end{equation}

\noindent
with $\epsilon \ll \sigma \sim 1$.
Here we have labelled the fermions with a family index, instead of the
names $s$, $b$, $\mu$, and $\tau$, since the mass matrices in this case 
are far from diagonal. A crucial point to notice is that the matrix for 
the leptons, which we will denote by $L$, is the transpose of the 
matrix for the down quarks, which we will denote by $D$. This is a 
feature of minimal $SU(5)$. It arises from the fact that the 
$\overline{{\bf 5}}$ representation of fermions contains the 
left-handed leptons, $l_L$, and
the charge conjugate of the right-handed down-quarks, $d_R$, while
the ${\bf 10}$ representation of fermions contains the charge
conjugate of the right-handed leptons, $l_R$, and the 
left-handed down-quarks, $d_L$.  Thus, $SU(5)$ 
relates $D$ to $L$, but only up to a left-right transposition: $D = L^T$.

The transposition feature of $SU(5)$ unification appearing in Eq. (6)
results in the large element, $\sigma$, of $L$ producing 
an O(1) mixing of $l_{2L}$ with $l_{3L}$ for the leptons, while in $D$ 
for the quarks it produces a large mixing of the right-handed 
fields $d_{2R}$ and $d_{3R}$.  The mismatch between the large
$l_{2L}- l_{3L}$ mixing and  
the $\nu_{2L}-\nu_{3L}$ mixing, which is small (as will soon be seen),
leads to a 
large $U_{\mu 3}$ mixing element.  But the right-handed mixings of the 
quarks are not observable through standard model physics.  What 
matters is the left-handed mixing of $d_{2L}$ with $d_{3L}$, which 
contributes to $V_{cb}$, and is controlled by the small parameter 
$\epsilon$.

The common statement that grand unification relates quark and lepton
mixing angles, and thus $V_{cb}$ to $U_{\mu 3}$, is very misleading.
What is really true in general is that grand unification relates
the mixing of quarks of one handedness to the mixing of leptons of
the other handedness. Thus $V_{cb}$ and $U_{\mu 3}$ need not be
directly related to each other.  Of course, if the mass matrices are 
symmetric, as has almost always been assumed, the left-handed and right-handed 
mixings are the same, and hence $V_{cb}$ is directly related to 
$U_{\mu 3}$. The most natural interpretation, then, of the experimental 
discovery that $|U_{\mu 3}| \gg |V_{cb}|$ is that the mass matrices are 
highly asymmetric. This is the essential point first made in \cite{lop}.

Not only does a highly asymmetric, or, as we will call it, ``lopsided,"
form of the mass matrices explain the difference between the size
of $U_{\mu 3}$ and $V_{cb}$, but it also explains the fact, noted above,
that $U_{\mu 3}$ is larger than the Fritzsch value and $V_{cb}$ is
smaller than the Fritzsch value by about the same factor. The point
is that the product of the two off-diagonal elements, $\epsilon$
and $\sigma$, is controlled by the fermion mass ratio. As is evident
from Eq. (6), $m_s/m_b \cong \frac{\epsilon \sigma}{1 + 
\sigma^2} \sim \epsilon \sigma$. That means that
the Fritzsch prediction for the mixing of $d_L$ and $s_L$, which
is $\sqrt{m_s/m_b}$, goes approximately as $\sqrt{\epsilon
\sigma}$. That shows that the Fritzsch prediction for the mixing angles  
is roughly the geometric mean between the true value of $U_{\mu 3} 
\sim \sigma$ and the true value of $V_{cb}$ $\sim \epsilon$. In other 
words, in our hypothesis of lopsided mass matrices, the surprising 
largeness of $U_{\mu 3}$ and the surprising smallness of $V_{cb}$ are 
two sides of the same coin.

Another important feature of this mechanism should be emphasized. Almost
all published explanations of the largeness of the $\nu_{\mu}-\nu_{\tau}$
mixing trace it to some special feature or form of the neutrino
mass matrix. Perhaps this is due to the purely linguistic fact that
we talk about ``neutrino mixing angles". But they could just as well
be called the ``charged-lepton mixing angles". They are really the angles
expressing the mismatch between the neutrino mass eigenstates and the
charged lepton mass eigenstates, just as the CKM angles are the mismatch
between the up and down quark eigenstates. In our mechanism, the large 
value of $U_{\mu 3}$ is traceable to a peculiarity of the charged 
lepton mass matrix $L$, namely, having a large off-diagonal entry 
$\sigma$. As we shall see in the next Section, having such a large 
entry helps to explain several other features of the quark and lepton 
mass spectrum. 

To sum up, the mechanism for explaining large $\nu_{\mu}-\nu_{\tau}$
mixing proposed in \cite{ab1,abb,ab4} has three salient features:
(1) the largeness of this mixing is due to the charged lepton
mass matrix, which is (2) highly asymmetric, and which is (3) related
to the transpose of the down quark mass matrix by $SU(5)$.

In the next Section we will see how a model with precisely these 
features arises very naturally in $SO(10)$ from very different 
considerations.

\section{FERMION MASS MATRICES IN MINIMAL SCHEMES OF $SO(10)$ BREAKING}

The model that we shall examine in this paper emerged originally
from our attempt to construct a realistic model based on 
$SO(10)$ \cite{so10} in which
$SO(10)$ is broken to the standard model group, $G_{SM} =
SU(3)_c \times SU(2)_L \times U(1)_Y$ in the simplest possible
way. We shall therefore start by explaining what we mean by
minimal $SO(10)$ breaking.

Since $SO(10)$ is a rank 5 group, it requires for its breakdown to 
$G_{SM}$ at least two Higgs fields. One Higgs field is needed to
break the rank of the group to 4, but this generally leaves 
an unbroken $SU(5)$. The second Higgs field is needed to break $SU(5)$
down to $G_{SM}$. The two breakings can occur in either sequence 
depending upon which Higgs field has the larger VEV and effects the 
first breaking.

Whatever Higgs field gives superlarge mass to the right-handed neutrinos,
as required for the standard seesaw explanation of the lightness of
the left-handed neutrinos, will also break $SO(10)$ down to $SU(5)$,
and thus the rank to 4. There are two simple choices for this Higgs 
field: either an antisymmetric five-index tensor $\overline{{\bf 126}}$
or a spinor $\overline{{\bf 16}}$.  In either case,
one also expects a Higgs field in the conjugate representation,
${\bf 126}$ or ${\bf 16}$, to go along with it.  A nice feature of
the $\overline{{\bf 126}}$ is that this tensorial representation 
leaves unbroken a $Z_2$ subgroup of the center of $SO(10)$ that acts as
an automatic matter parity, whereas if a spinor Higgs is introduced,
then matter parity is not automatic. On the
other hand, to introduce ${\bf 126} + \overline{{\bf 126}}$ 
is to introduce quite large representations that
tend to make the unified gauge coupling go non-perturbative below the 
Planck scale, and that may be hard to obtain
from superstring theory. In any event, it would seem that the use
of a spinor-antispinor pair, ${\bf 16} + \overline{{\bf 16}}$, is
more economical. Thus we assume that the rank of $SO(10)$ is broken at
the unification scale and the right-handed neutrinos get mass from
one such spinor-antispinor pair of Higgs fields.

To break the group the rest of the way to $G_{SM}$ requires the existence
of Higgs fields in the adjoint representation ${\bf 45}$ and/or in the 
symmetric two-index tensor representation ${\bf 54}$. Most published 
realistic $SO(10)$ models have several of both kinds of multiplets.
However, it has been shown that it is possible to break 
$SO(10)$ to $G_{SM}$ with only a {\it single} adjoint Higgs and no 
larger representations \cite{b-r}.

This, then, is what we call the ``minimal breaking scheme for $SO(10)$": 
{\it The breaking of $SO(10)$ to $G_{SM}$ is accomplished by the 
expectation values of a set of Higgs fields consisting of} ${\bf 45}_H 
+ {\bf 16}_H + \overline{{\bf 16}}_H$, {\it with the model containing 
no multiplets larger than the single} ${\bf 45}$. There, of course, have to 
be other Higgs fields to break the $SU(2)_L \times U(1)_Y$ group of 
the electroweak interactions.

This minimality assumption is restrictive enough that it is possible to
say in which direction the expectation values of these fields point.
This can be done by considering the problem of doublet-triplet splitting,
whereby the colored partners of the weak-doublet Higgs
fields of the standard model become superheavy while the weak-scale
masses of the doublets themselves are preserved. In $SO(10)$ the only 
known way of doing this in a technically natural manner is the 
Dimopoulos-Wilczek or ``missing VEV" mechanism \cite{dim-wil}. The idea 
is that if an adjoint Higgs field
that has an expectation value proportional to the $SO(10)$ generator
$B-L$ couples to Higgs fields in the vector representation, it will 
make their color-triplet components heavy (since they have $B-L = 
\pm 2/3$) while leaving their weak-doublet components massless 
(since they have $B-L = 0$).
The needed coupling is simply of the form ${\bf 10}_{1H} {\bf 45}_H
{\bf 10}_{2H}$. Of course, the expectation value of the adjoint, by
virtue of the definition of the adjoint representation, is necessarily
a linear combination of generators of the group. $B-L$ is one of the
$SO(10)$ generators that is picked out by simple forms of the Higgs 
superpotential for the adjoint multiplet. It should be noted that
there is another version of the missing VEV mechanism that works in 
$SO(10)$, in which the VEV of the adjoint is proportional to the 
generator $I_{3R}$ of the $SU(2)_R$ subgroup of $SO(10)$ \cite{I3R}. 
However, that version is significantly more complicated. Therefore, 
simplicity dictates the choice that

\begin{equation}
\langle {\bf 45}_H \rangle \propto B-L.
\end{equation}

\noindent
Since the assumption of a minimal $SO(10)$ breaking scheme included
the supposition that only {\it one} adjoint exists in the model, no 
adjoint exists except the one that points in the $B-L$ direction. As 
we shall see, this puts an important limitation on the possibilities for 
constructing realistic mass matrices for the quarks and leptons.
The assumption of a minimal $SO(10)$ breaking scheme thus acts as
an important guide in searching for good models.

The simplest possible terms that would give mass to quarks and leptons
in $SO(10)$ would be $\lambda_{ij} {\bf 16}_i {\bf 16}_j {\bf 10}_H$, 
where the subscripts $i$ and $j$ are family indices.
This would lead to four proportional Dirac mass matrices 
for the up-quarks ($U$), down quarks ($D$), charged leptons ($L$),
and neutrinos ($N$). In fact one would have $D = L 
\propto U = N$. Moreover, all these matrices would be symmetric, which 
is why one can write $D= L$ instead of $D = L^T$ as in minimal $SU(5)$. 
Some of the predictions that follow from these relations are good, 
notably the famous 
prediction $m_b^0 = m_{\tau}^0$, where the superscript zero stands for 
quantities evaluated at the unification scale $M_G$. However, $D=L$ 
also predicts that $m_s^0 = m_{\mu}^0$ and $m_d^0 = m_e^0$. Empirically,
one finds instead that $m_s^0 \simeq \frac{1}{3} m_{\mu}^0$ and 
$m_d^0 \simeq 3 m_e^0$. These factors of three are called the 
Georgi-Jarlskog factors \cite{g-j}. The simplest possible $SO(10)$ 
Yukawa terms also predict that all the CKM angles vanish, since $U 
\propto D$. While not exactly true, this is at least a good zeroth 
order relation, since the CKM angles are all small compared to unity. 
By contrast, in $SU(5)$ the matrices $D$ and $U$ are not related by 
the unified symmetry and so the CKM angles are unconstrained. The 
smallness of the CKM angles can be regarded, therefore, as evidence 
for $SO(10)$. On the other hand, the proportionality of $D$ and $U$ 
in $SO(10)$ also predicts that $m_c^0/m_t^0 = m_s^0/m_b^0$, which
fails badly by over an order of magnitude.

What one can conclude is that a way of going beyond the simplest possible
$SO(10)$ Yukawa scheme must be found which preserves some of its
predictions while breaking others. One way to do this involves using
larger representations to break the electroweak interactions. For 
instance, in the original Georgi-Jarlskog model, a 
$\overline{{\bf 45}}$ multiplet of $SU(5)$ (not to be confused 
with the adjoint of $SO(10)$) participates in breaking $SU(2)_L 
\times U(1)_Y$. In the context of $SO(10)$, this $\overline{{\bf 45}}$ 
is contained in a $\overline{{\bf 126}}$, which is inconsistent 
with our minimality assumptions. More economical is to assume 
that the Higgs fields that break $SO(10)$ at the unification scale,
i.e., the ${\bf 45}_H + {\bf 16}_H + \overline{{\bf 16}}_H$, couple
to quarks and leptons and thus introduce the effects of that $SO(10)$ 
breaking into the quark and lepton mass relations. This is the 
assumption we make.

To describe the third family it is simplest to assume the minimal Yukawa
term ${\bf 16}_3 {\bf 16}_3 {\bf 10}_H$ as pictured in Fig. 1(a).
By itself, this would make 
all the mass matrices have the form 

\begin{equation}
\left( \begin{array}{ccc} 0 & 0 & 0 \\ 0 & 0 & 0 \\ 0 & 0 & 1
\end{array} \right).
\end{equation}

\noindent
That would give the following predictions, all of which are at least
good zeroth approximations to reality: $m_b^0 = m_{\tau}^0$, $V_{cb} 
= 0$, and $m_1/m_3 = m_2/m_3 =0$, where $m_i$ is a mass of a fermion 
of the $i^{th}$ family. Note that these are just the ``good" $SO(10)$ 
predictions mentioned above.

The second family presents more of a challenge. The main issue is how 
to get the Georgi-Jarlskog factor of 3 between $m_{\mu}^0$ and $m_s^0$. 
Breaking of $SU(5)$ must be involved, since the bad relation
$m_s^0 = m_{\mu}^0$ arises already at the $SU(5)$ level. The only 
field that breaks $SU(5)$ in the framework of minimal $SO(10)$ breaking 
is the adjoint, ${\bf 45}_H$. Since $\langle {\bf 45}_H \rangle \propto
B-L$, and the $B-L$ of leptons is $-3$ times that of quarks, this field
has the possibility of giving the needed Georgi-Jarlskog factor. Thus 
one must seek an effective Yukawa term that involves the ${\bf 45}_H$. 
The simplest such term \cite{adjointterm}, 
in the sense of the term of lowest dimension, 
is of the form $({\bf 16}_i {\bf 16}_j) {\bf 10}_H {\bf 45}_H/M_G$. Moreover,
this term can arise in a simple way by the integration out of a ${\bf 16} + 
\overline{{\bf 16}}$ family plus antifamily at the unification scale,
as shown in Fig. 1(b).

There are actually two ways to contract the $SO(10)$ indices of such a 
term: the product $({\bf 16}_i {\bf 16}_j)$ can be contracted 
symmetrically or antisymmetrically. It is easy to show that if 
$\langle {\bf 45} \rangle \propto B-L$, only the antisymmetric contraction
contributes to the quark and lepton mass matrices. (The reason is
simple. If the VEV of the adjoint is proportional to a generator $Q$,
then the symmetric/antisymmetric contractions give contributions
to fermion masses that go as $Q(f) \pm Q(\overline{f})$. Since $B-L$
of an antifermion is minus that of the fermion, the
contribution cancels for the symmetric contraction.) Thus, one need only
consider the flavor-antisymmetric term, which means only
$ij = 23$ and not $ij = 22$, or $33$.  Consequently, the only operator of
interest is $({\bf 16}_2 {\bf 16}_3) {\bf 10}_H {\bf 45}_H$ which, together 
with the operator ${\bf 16}_3 {\bf 16}_3 {\bf 10}_H$, gives

\begin{equation}
\begin{array}{ll} 
U = \left( \begin{array}{ccc} 
0 & 0 & 0 \\ 0 & 0 & \epsilon/3 \\ 0 & - \epsilon/3 & 1 \end{array}
\right) M_U, \;\;\; & D = \left( \begin{array}{ccc}
0 & 0 & 0 \\ 0 & 0 & \epsilon/3 \\ 0 & - \epsilon/3 & 1 \end{array}
\right) M_D, \\ & \\
N = \left( \begin{array}{ccc}
0 & 0 & 0 \\ 0 & 0 & - \epsilon \\ 0 & \epsilon & 1 \end{array}
\right) M_U, \;\;\; & L = \left( \begin{array}{ccc}
0 & 0 & 0 \\ 0 & 0 & -\epsilon \\ 0 & \epsilon & 1 \end{array}
\right) M_D. \\ &  
\end{array}
\end{equation}

The desired factor of 3 has been achieved between leptons and quarks, due
to the generator $B-L$ to which the adjoint VEV is proportional. One 
also can see that the $\epsilon$ entries are flavor antisymmetric for 
reasons already explained. As they stand, these forms of the matrices 
are inadequate to explain even the features of the second and third 
families of fermions. There are three inadequacies. (1) The factor of 
3 comes in squared between the mass of the leptons and quarks 
of the second family.  The reason is that, for $\epsilon$ small due to 
the mass hierarchy between families, the second eigenvalue of 
$L$ is given by the seesaw formula $m_{\mu}^0 \cong \epsilon^2 M_D$, 
while the second eigenvalue of $D$ is given by $m_s^0 \cong 
(\epsilon/3)^2 M_D$. (2) The matrices $D$ and $U$ are still exactly 
proportional. This is a 
consequence of the fact that the generator $B-L$ does not distinguish
up and down quarks. Therefore, the CKM angle $V_{cb}$ still exactly
vanishes. (3) Because $D$ and $U$ are exactly proportional,
one still has the bad prediction $m_c^0/m_t^0 = m_s^0/m_b^0$. 

It is clear that the breaking of $SO(10)$ due to the adjoint cannot cure
all of these problems, since $B-L$ does not distinguish
$D$ from $U$. Thus the breaking of $SO(10)$ done by
${\bf 16}_H + \overline{{\bf 16}}_H$ must come into play. As we shall 
now show, a single simple operator exists, which involves one of these 
spinor Higgs and cures at one stroke all three of the problems we have 
identified.

The lowest dimension effective Yukawa operators that involve the spinor
Higgs fields are quartic in spinors. Consider, therefore, operators of
the form ${\bf 16}_i {\bf 16}_j {\bf 16'}_H {\bf 16}_H/M_G$. 
The ${\bf 16}_H$ is the spinor Higgs field that breaks $SO(10)$ at $M_G$
down to $SU(5)$. The ${\bf 16'}_H$ is a spinor Higgs that has a 
weak-scale VEV that breaks $SU(2)_L \times U(1)_Y$. In principle, 
these two spinors could be the same field. However, if they were, 
it would mean that they had to be contracted symmetrically by Bose 
statistics, which in turn would mean that ${\bf 16}_i$ and ${\bf 16}_j$
would also have to be contracted symmetrically. A careful examination 
shows that the resulting flavor-symmetric contributions to the mass 
matrices do not lead to realistic forms, though it is possible to achieve
realistic mass matrices by adding yet another Yukawa operator, as in
the interesting model of Babu, Pati, and Wilczek \cite{bpw}.
Therefore ${\bf 16'_H}$ must be a distinct field. 
As will be seen later, introducing this ${\bf 16}'_H$ involves no loss
of economy, since it allows a very elegant explanation of the largeness
of the ratio $m_t/m_b$ without making $\tan \beta$ large.

There are still several operators of this type to be considered: the 
family indices can take the values $ij = 33, 22, 23$, or $32$, and 
there are three ways to contract the four spinors to make an $SO(10)$ 
singlet. Here again, one must examine the various cases to see which 
gives the most realistic mass matrices. As it turns out, there is one 
operator that is much superior to the others, in the sense that it 
much more cleanly and simply fits the data. It is of the form 
$[{\bf 16}_2 {\bf 16}_H] [{\bf 16}_3 {\bf 16'}_H]$, where $[...]$ 
means that the spinors inside are contracted into a ${\bf 10}$. 
This can arise very simply by integrating out a ${\bf 10}$ of 
fermions, as shown in Fig. 1(c).

Let us write the resulting mass operator in $SU(5)$ language. Denote by
${\bf p}({\bf q})$ a ${\bf p}$ multiplet of $SU(5)$ that is contained
in a ${\bf q}$ multiplet of $SO(10)$. The VEV
of ${\bf 16}_H$ lies, of course, in the ${\bf 1}({\bf 16})$ direction,
while the VEV of ${\bf 16}'_H$ that breaks the weak interactions lies
in the $\overline{{\bf 5}}({\bf 16})$ direction. Therefore, the
resulting mass term is of the form $[\overline{{\bf 5}}({\bf 16}_2) 
{\bf 1}({\bf 16}_H)] [{\bf 10}({\bf 16}_3) \overline{{\bf 5}} 
({\bf 16}'_H)]$, which in $SU(5)$ terms gives effectively the operator 
$(\overline{{\bf 5}}_2 {\bf 10}_3) \overline{{\bf 5}}_H$. 
Note that this has the same form as the $SU(5)$ operator discussed
in the last Section, which gave the $\sigma$ entries in Eq. (6). The 
result, then, of including this operator \cite{abb} is to make the 
mass matrices
take the form:

\begin{equation}
\begin{array}{ll}
U = \left( \begin{array}{ccc} 0 & 0 & 0 \\
0 & 0 & \epsilon/3 \\ 0 & - \epsilon/3 & 1 \end{array} \right) M_U, 
\;\;\; & 
D = \left( \begin{array}{ccc} 0 & 0 & 0 \\
0 & 0 & \sigma + \epsilon/3 \\ 0 & - \epsilon/3 & 1 \\ \end{array}
\right) M_D, \\ & \\
N = \left( \begin{array}{ccc} 0 & 0 & 0 \\
0 & 0 & - \epsilon \\ 0 & \epsilon & 1 \end{array} \right) M_U, \;\;\; &
L = \left( \begin{array}{ccc} 0 & 0 & 0 \\
0 & 0 & - \epsilon \\ 0 & \sigma + \epsilon & 1 \end{array} \right) 
M_D \\ & \end{array}.
\end{equation}

\noindent
The new term has given the entries we call $\sigma$. Note that
these lopsided entries appear only in $D$ and $L$. The reason is 
simply that the ${\bf 16}'_H$ contains a $\overline{{\bf 5}}$ of 
$SU(5)$ but no ${\bf 5}$.

It is easy to see that the new term with $\sigma \gg \epsilon$ cures at 
once all three of the problems we identified with the forms given in 
Eq. (9):
(1) Instead of $m_{\mu}^0 \cong \epsilon^2 M_D$ and 
$m_s^0 \cong (\epsilon/3)^2 M_D$, one has approximately that $m_{\mu}^0 
\propto (\epsilon)(\sigma + \epsilon) \cong \epsilon \sigma$ and $m_s^0  
\propto (\epsilon/3)(\sigma + \epsilon/3) \cong \epsilon \sigma/3$. 
More exact expressions will be given later. Thus the desired
Georgi-Jarlskog factor of 1/3 is obtained, instead of 1/9. The
$\sigma$ entry has dominated over one of the factors of 
$\epsilon/3$ and thus prevented the factor of 1/3 from coming in squared.

(2) The $\sigma$ entry comes into $D$ but not $U$, and thus breaks
the proportionality of the two matrices. As a result, $V_{cb}$ no
longer vanishes, but is given approximately by $(\epsilon/3)
(\frac{\sigma^2}{\sigma^2 + 1})$. Note that this is of the same
order in $\epsilon$ as $m_s/m_b \cong (\epsilon/3)(\frac{\sigma}
{\sigma^2  + 1})$, rather than $\sqrt{m_s/m_b}$ as
is the case with Fritzsch forms, and accords much better with the actual 
experimental values.

(3) The fact that $\sigma$ breaks the proportionality of $U$ and $D$
also means that the bad relation $m_c^0/m_t^0 = m_s^0/m_b^0$ is
broken.  Specifically, $m_s^0/m_b^0$ is of order
$\epsilon$, while $m_c^0/m_t^0$ is still of order $\epsilon^2$ and
therefore much smaller. This also accords well with the experimental 
numbers. In fact, as we shall see, if one uses $V_{cb}$ and 
$m_{\mu}/m_{\tau}$ to fix the parameters $\sigma$ and $\epsilon$, 
one finds that $m_c$(1 GeV) is predicted to be in agreement with the  
experimentally determined value of $1.27 \pm 0.1$ GeV. 
It should also be noted that the prediction $m_b^0 = m_{\tau}^0$
is only very slightly affected by the addition of the $\sigma$ term,
both $m_b^0$ and $m_{\tau}^0$ being given to leading order in $\epsilon$
by $\sqrt{\sigma^2 + 1} M_D$.

The economy of the above mass matrix forms is seen in the fact that
five quantities ($V_{cb}$, $m_{\mu}/m_{\tau}$, $m_s/m_b$, $m_c/m_t$, and
$m_{\tau}/m_b$) are successfully fit with only the two parameters
$\sigma$ and $\epsilon$. No other published form succeeds in accurately
reproducing the masses and mixing of the the heavier two families with so
few parameters. The predictions and fits will be discussed in detail in
Section IV.

We see that the matrices in Eq. (10) were arrived at by a process of
reasoning that had nothing to do with the question of neutrino mass
but rather with an attempt to get
realistic masses and mixings for the quarks and
charged leptons using as simple a Higgs sector as possible in
$SO(10)$. But what has emerged is a structure with precisely the
three critical features identified in the last Section as giving a
simple explanation of the large mixing of $\nu_{\mu}$ and $\nu_{\tau}$.
In fact, a fit of $V_{cb}$ and $m_{\mu}/m_{\tau}$ gives $\sigma
\cong 1.8$ and $\epsilon \cong 0.14$. Consequently, as can be seen
directly from Eq. (10), the angle $\theta_{\mu \tau} = \sin^{-1} U_{\mu 3}
= \tan^{-1} \sigma - O(\epsilon) = 60^{\circ} - O(8^{\circ})$. This
is quite consistent with what is observed. We will look more carefully
at these predictions later.

To summarize, with two parameters, $\epsilon$ and $\sigma$, four mass
ratios and two mixing angles are satisfactorily accounted for, if we 
include $U_{\mu 3}$. No greater economy could be hoped for in explaining
the spectrum of the heavy two families. Moreover, as we shall see in the
next Section, the forms in Eq. (10) can be extended to include the first 
family with equal economy: the introduction of two new parameters 
(one of which is complex) will nicely account for seven quantities 
pertaining to the first family.

Before explaining how the model is extended to the first family
we will expand on a couple of points made earlier. First, we said that
the introduction of the ${\bf 16}'_H$ allows a simple explanation 
\cite{ab1,bpw} of
why $m_t \gg m_b$ that does not require a $\tan \beta \gg 1$.
The point is that the Higgs doublet of the MSSM that is often called
$H_U$ is purely contained in the ${\bf 10}_H$ that couples to
${\bf 16}_3 {\bf 16}_3$ and gives rise to the ``1" entry in the
mass matrices of Eq. (10). However, the Higgs doublet of the MSSM that
is called $H_D$ does not come purely from ${\bf 10}_H$. Rather it is a 
mixture of doublets in ${\bf 10}_H$ and ${\bf 16}'_H$, since they
both contain $\overline{{\bf 5}}$'s of $SU(5)$. Thus we may write

\begin{equation}
\begin{array}{l}
H_U = H({\bf 10}_H), \\ \\
H_D = \overline{H} ({\bf 10}_H)\cos \gamma  +
\overline{H}({\bf 16}'_H)\sin \gamma , 
\end{array}
\end{equation}

\noindent
where $\gamma$ is some 
mixing angle that depends on the parameters of the Higgs sector.
Since the 33 elements of the mass matrices all arise purely from the
coupling of the ${\bf 10}_H$, the parameters we called $M_U$ and $M_D$
in Eq. (10) are given by

\begin{equation}
\begin{array}{l}
M_U = \lambda_{33} \langle H_U \rangle, \\ \\ 
M_D = \lambda_{33} \langle H_D \rangle \cos \gamma.
\end{array}
\end{equation}

\noindent
leading to the ratio $M_U/M_D = \langle H_U \rangle/
(\langle H_D \rangle \cos \gamma) = \tan \beta/\cos \gamma$, where 
$\tan \beta$ is defined to be the ratio of the Higgs VEV $\langle H_U \rangle$
giving mass to the top quark to the Higgs VEV $\langle H_D \rangle$ giving
mass to the bottom quark. Hence

\begin{equation}
m_t^0/m_b^0 \cong (\sigma^2 + 1)^{-1/2} (\tan \beta/\cos \gamma)
\end{equation}

\noindent
The point is simply that the large ratio of the top to bottom masses
could be the result of $\cos \gamma$ being small rather than 
$\tan \beta$ being large. In fact, since we do not know anything
{\it a priori} about the angle $\gamma$, we cannot say whether $\tan 
\beta$ is large or small. It should be noted that if one assumes
that $H_D$ lies mostly in the ${\bf 16}'_H$ (so that $\cos \gamma \ll 
1$), it would explain why the parameter $\sigma$ is large since it 
comes from a coupling to ${\bf 16}'_H$, and also explain why $\tan 
\beta$ might be small.

A second point we wish to underline here has to do with the 
reasonableness of asymmetric mass matrices. In many models it is 
assumed that all the
mass matrices are symmetric. However, this is not something that is
called for by the group theory of grand unification. It is true that 
with the minimal Yukawa terms $SU(5)$ gives a symmetric $U$. But 
even with minimal Yukawa terms $SU(5)$ does not predict any symmetry
of the $D$ and $L$ matrices. And in $SO(10)$, as we have seen, once
one introduces the effects of $SO(10)$ breaking into the Yukawa sector,
as one virtually must, one easily obtains effective Yukawa terms that
are asymmetric. Fig. 1(c) shows that very simple diagrams can give terms 
that are lopsided, in the sense that they contribute only above or 
below the diagonal. From the point of view of the fundamental grand 
unified theory, then, lopsided terms are as natural as symmetric ones. 
The preference for symmetric terms has been the result not of 
examining what kinds of terms are obtained in a simple way in 
unification, but rather from the
desire to reduce the number of parameters at the level of mass matrices
with the aim of making models which are highly predictive.
However, putting oneself in the straightjacket of symmetric matrices
makes it hard to get a good fit to all the quark and lepton masses and
mixings. It turns out, as we have seen, and will see further below, that
allowing asymmetric matrices makes possible a model which gives both
a very good fit to the data and is actually much more predictive than 
most models which assume symmetric matrices.

\section{EXTENSION to the FIRST FAMILY}

In arriving at the form of the mass matrices for the heavy two families 
we were limited in the choices that were possible by the assumption we
made about the simplicity of the $SO(10)$-breaking sector. In 
extending to the first family we are not quite so limited. Nevertheless,
the number of simple possibilities is not very large. There are several
discrete choices: Should the contributions to the first row and column
of the mass matrices be flavor symmetric like the 1's in Eq. 10,
antisymmetric like the $\epsilon$'s, or lopsided like the $\sigma$'s?
Should they contribute to all the matrices equally like the 1's,
to all the matrices but with non-trivial Clebsch factors like the 
$\epsilon$'s or only to $D$ and $L$ like the $\sigma$'s? 
It is fairly easy to run through the various cases and see what kinds
of relations among masses and mixings result. As it turns out, one
of the simplest possibilities gives a remarkably good fit to the data. 
This uniquely simple choice \cite{ab4} is the following:

\begin{equation}
\begin{array}{ll}
U = \left( \begin{array}{ccc} \eta & 0 & 0 \\ 0 & 0 & \epsilon/3 \\
0 & - \epsilon/3 & 1 \end{array} \right) M_U, \;\;\; & 
D = \left( \begin{array}{ccc} \eta & \delta & \delta' \\
\delta & 0 & \sigma + \epsilon/3 \\ \delta' & - \epsilon/3 & 1 
\end{array} \right) M_D, \\ & \\
N = \left( \begin{array}{ccc} \eta & 0 & 0 \\ 0 & 0 & - \epsilon \\
0 & \epsilon & 1 \end{array} \right) M_U, \;\;\; & 
L = \left( \begin{array}{ccc} \eta & \delta & \delta' \\
\delta & 0 & -\epsilon \\ \delta' & \sigma + \epsilon & 1 \end{array}
\right) M_D.
\end{array}
\end{equation}

\noindent
We have already mentioned that fits give $\sigma \cong 1.8$ and 
$\epsilon \cong 0.14$. The new parameters $\delta$ and $\delta'$ both 
have magnitude of about $0.008$. The parameter $\eta$ is by far the 
smallest, being about $8 \times 10^{-6}$. The only role that $\eta$ 
plays in the sector of quarks and charged leptons is in giving the 
up quark a mass, for it makes negligible contributions to the down
quark and electron masses as determined from $D$ and $L$, respectively.
In Fig. 2(a) we have illustrated a higher-order diagram that can 
contribute to the parameter $\eta$.
Since it is not excluded that the up quark is exactly massless, it is 
possible to set $\eta$ to zero. In any event, one can see that $\eta 
\cong m_u^0/m_t^0$, which is by orders of magnitude smaller than any 
other interfamily ratio of masses in the standard model. It will, 
however, be of some significance for neutrino masses. 
If $\eta$ vanishes, this model gives only the small-angle MSW solution 
to the solar neutrino problem. But even if $\eta$ is as small as
$8 \times 10^{-6}$, it allows either the small-angle MSW solution 
or bimaximal neutrino mixing to arise in a simple way.

Turning to the parameters $\delta$ and $\delta'$, we see that they
appear symmetrically and only in $D$ and $L$. 
Such terms are easily obtained in $SO(10)$ from simple diagrams
such as that shown in Figs. 2(b) and 2(c). The effective operators 
arising from
these diagrams are of the form $[{\bf 16}_1 {\bf 16}_j][{\bf 16}_H
{\bf 16}'_H]$ with $j = 2,3$, where again the 
spinors in brackets are contracted symmetrically into a ${\bf 10}$ 
of $SO(10)$ which is integrated out.  Note, however, that the 
symmetric contributions $\delta$ and $\delta'$ from the two Higgs
contraction $[{\bf 1}({\bf 16}_H){\overline{\bf 5}}({\bf 16}'_H)]$
contributes only to $D$ and $L$ by virtue of their $SU(5)$ 
structure.  Contrast these effective operators for $j = 2,3$ with that
occurring previously for the term $\sigma$ arising from the diagram
shown in Fig. 1(c).  

The three new parameters we have introduced are, as we shall see, 
sufficient to account for everything about the first family.
Before proceeding, however, we must be careful about complex phases. 
It is easy to show that if we allow all parameters of the model to be 
complex all but two phase angles can be rotated away from the mass 
matrices $U$, $D$, $L$ and $N$, provided we now neglect the 
negligible $\eta$ contributions to $D$ and $L$.  We will call these 
physical phases $\alpha$ and $\phi$ which appear as follows, 

\begin{equation}
\begin{array}{ll}
U = \left( \begin{array}{ccc} \eta & 0 & 0 \\
  0 & 0 & \epsilon/3 \\ 0 & - \epsilon/3 & 1 \end{array} \right) M_U, 
  \;\;\; &
D = \left( \begin{array}{ccc} 0 & \delta & \delta' e^{i (\phi + \alpha)}
  \\
  \delta & 0 & \sigma + \epsilon e^{i \alpha}/3  \\
  \delta' e^{i \phi} & - \epsilon/3 & 1 \end{array} \right) M_D, \\ & \\
N = \left( \begin{array}{ccc} \eta & 0 & 0 \\ 0 & 0 & - \epsilon \\
  0 & \epsilon e^{i \alpha} & 1 \end{array} \right) M_U, \;\;\; &
L = \left( \begin{array}{ccc} 0 & \delta & \delta' e^{i \phi} \\
  \delta & 0 & -\epsilon \\ \delta' e^{i (\phi + \alpha)} & 
  \sigma + \epsilon e^{i \alpha} & 1 \end{array} \right) M_D,
\end{array}
\end{equation}

\noindent
where in these matrices and henceforth $\epsilon$, $\sigma$, $\delta$,
$\delta'$ and $\eta$ denote the magnitudes of these parameters and 
the phases are written explicitly. The phase $\alpha$ only comes into
the fits of masses at higher order in the small quantity 
$\epsilon/\sigma \cong 0.08$.
Numerically, its effect is only a few percent; moreover, the fits 
(especially to $m_c$) prefer a value near zero. Therefore, we can ignore
$\alpha$ and will do so from now on. That leaves only the phase $\phi$. 
Its only significant effect, but a very important one, is to
give the CP-violating phase angle $\delta_{CP}$. 

Instead of using the parameters $\delta$, $\delta'$ and $e^{i \phi}$,
it will be somewhat more convenient to use the parameters $t_L$, 
$t_R$, and $e^{i \theta}$, which are defined in terms of them as follows:

\begin{equation}
t_L e^{i \theta} \equiv \frac{\delta - \sigma \delta' e^{i \phi}} 
{\sigma \epsilon/3},
\end{equation}

\noindent
and

\begin{equation}
t_R \equiv \frac{\delta \sqrt{\sigma^2 + 1}}{\sigma \epsilon/3}.
\end{equation}

\noindent
The significance of these parameters is that they are essentially
the left-handed and right-handed Cabbibo angles. This can be seen
by taking the forms for $D$ and $L$ given in Eq. (15) and diagonalizing
the 2-3 block. When this is done the 1-2 blocks of these matrices take
the form 

\begin{equation}
D^{[12]} \propto \left( \begin{array}{cc}
0 & t_R \\ t_L & 1 \end{array} \right), \;\;\;\; L^{[12]}
\propto \left( \begin{array}{cc} 0 & t_L \\ t_R & 3 \end{array}
\right). 
\end{equation}

In terms of the five dimensionless parameters $\epsilon$, $\sigma$,
$t_L$, $t_R$, and $e^{i \theta}$ with $\eta$ set equal to zero, we now write 
down expressions for fourteen observable quantities: seven ratios of quark and
lepton masses, three CKM angles and one phase, and three lepton mixing angles.

\begin{equation}
\begin{array}{lcl}
m_b^0/m_{\tau}^0 & \cong & 1 - \frac{2}{3} \frac{\sigma}{\sigma^2 + 1} 
\epsilon, \\[0.2in]
m_c^0/m_t^0 & \cong & \frac{1}{9} \epsilon^2 \cdot [1 - \frac{2}{9} 
\epsilon^2], \\[0.2in]
m_{\mu}^0/m_{\tau}^0 & \cong & \epsilon \frac{\sigma}{\sigma^2 + 1} \cdot
[1 + \epsilon \frac{1 - \sigma^2 - \sigma \epsilon}{\sigma(\sigma^2 + 1)}
+ \frac{1}{18}(t_L^2 + t_R^2)], \\[0.2in]
m_s^0/m_b^0 & \cong & \frac{1}{3} \epsilon \frac{\sigma}{\sigma^2 + 1}
\cdot [1  + \frac{1}{3} \epsilon \frac{1 - \sigma^2 - \sigma \epsilon/3}
{\sigma (\sigma^2 + 1)} + \frac{1}{2} (t_L^2 + t_R^2) ], \\[0.2in]
m_u^0/m_t^0 & = & 0, \\[0.2in] 
m_e^0/m_{\mu}^0 & \cong & \frac{1}{9} t_L t_R \cdot [1 - \epsilon 
\frac{\sigma^2 + 2}{\sigma(\sigma^2 + 1)} + \epsilon^2
\frac{\sigma^4 + 9 \sigma^2/2 + 3}{\sigma^2(\sigma^2 + 1)^2}
- \frac{1}{9} (t_L^2 + t_R^2)], \\[0.2in]
m_d^0/m_s^0 & \cong & t_L t_R \cdot [ 1 - \frac{1}{3} \epsilon
\frac{\sigma^2 + 2}{\sigma (\sigma^2 + 1)} - (t_L^2 + t_R^2)
+ (t_L^4 + t_L^2 t_R^2 + t_R^4)], \\[0.2in]
V_{cb} & \cong & \frac{1}{3} \epsilon \frac{\sigma^2}{\sigma^2 + 1}
\cdot [1 + \frac{2}{3} \epsilon \frac{1}{\sigma(\sigma^2 + 1)}], 
\\[0.2in]
V_{us} & \cong & t_L [ 1 -\frac{1}{2} t_L^2 - t_R^2 + t_R^4 + \frac{5}{2}
t_L^2 t_R^2 + \frac{3}{8} t_L^4 - \frac{\epsilon}{3 \sigma 
\sqrt{\sigma^2 + 1}} \frac{t_R}{t_L} e^{- i \theta}], \\[0.2in]
V_{ub} & \cong & \frac{1}{3} t_L \epsilon \frac{1}{\sigma^2 + 1} 
[\sqrt{\sigma^2 + 1} \frac{t_R}{t_L} e^{-i \theta} (1 - \frac{1}{3}
\epsilon \frac{\sigma}{\sigma^2 + 1}) - (1 - \frac{2}{3}
\epsilon \frac{\sigma}{\sigma^2 + 1})], \\[0.2in]
U_{\mu 3}^0 & \equiv & \sin \theta_{\mu \tau} 
\cong \frac{\sigma}{\sqrt{\sigma^2 + 1}} + O(\epsilon), \\[0.2in] 
U_{e 2}^0 & \cong & \cos \theta_{\mu \tau} \left( \frac{1}{3} t_R \right)
\cdot [ 1 + \epsilon (\frac{\tan \theta_{\mu \tau}}{\sigma^2 + 1}
- \frac{t_L}{t_R} e^{i \theta} \frac{(1 + \sigma \tan \theta_{\mu \tau})}
{\sigma \sqrt{\sigma^2 + 1}} ) - \frac{1}{18} t_R^2 - \frac{1}{9}
t_L^2 ], \\[0.2in]
U_{e 3} & \cong & \tan \theta_{\mu \tau} U_{e 2} \cdot [1 + \epsilon
\frac{2}{\sin 2 \theta_{\mu \tau}} \left( \frac{t_L}{t_R} e^{i \theta}
\frac{1}{\sqrt{\sigma^2 + 1}} - \frac{1}{\sigma^2 + 1} \right) ], \\ 
\end{array}
\end{equation}

\noindent
These expansions have been carried to sufficiently high order in small
quantities to be accurate to within 0.2\% and are useful 
in doing the fits to the data. However, the leading terms in these 
expansions have much simpler forms and thus allow one to see
more readily the relationships among various quantities in this model.
We therefore write these simpler expressions for purposes of discussion.

\begin{equation}
\begin{array}{lcl}
m_b^0/m_{\tau}^0 & \cong & 1, \\[0.2in]
m_c^0/m_t^0 & \cong & \frac{1}{9} \epsilon^2, \\[0.2in]
m_{\mu}^0/m_{\tau}^0 & \cong & \epsilon \frac{\sigma}{\sigma^2 + 1},
\\[0.2in]
m_s^0/m_b^0 & \cong & \frac{1}{3} \epsilon \frac{\sigma}{\sigma^2 + 1},
\\[0.2in]
m_u^0/m_t^0 & = & 0, \\[0.2in]
m_e^0/m_{\mu}^0 & \cong & \frac{1}{9} t_L t_R, \\[0.2in] 
m_d^0/m_s^0 & \cong & t_L t_R, \\[0.2in] 
V_{cb}^0 & \cong & \frac{1}{3} \epsilon \frac{\sigma^2}{\sigma^2 + 1},
\\[0.2in]
V_{us}^0 & \cong & t_L, \\[0.2in]
V_{ub}^0 & \cong & \frac{1}{3} t_L \epsilon \frac{1}{\sigma^2 + 1} 
(\sqrt{\sigma^2 + 1} \frac{t_R}{t_L} e^{-i \theta} - 1), \\[0.2in] 
U_{\mu 3}^0 & \equiv & \sin \theta_{\mu \tau} 
\cong \frac{\sigma}{\sqrt{\sigma^2 + 1}} + O(\epsilon) \sim 0.7, 
\\[0.2in]
U_{e2}^0 & \cong & \cos \theta_{\mu \tau} \left( \frac{1}{3} t_R \right),
\\[0.2in]
U_{e3} & \cong & \sin \theta_{\mu \tau} \left( \frac{1}{3} t_R \right).
\\ & &
\end{array}
\end{equation}

It might at first seem surprising that without any information about 
the Majorana mass matrix $M_R$ of the right-handed neutrinos 
we are able to write down predictions for the three neutrino mixing 
angles. However, if $\eta = 0$, as we are assuming at present, then
the Dirac mass matrix of the neutrinos ($N$) has vanishing first row 
and column, and therefore, obviously, the same will be true of
the mass matrix of the light
neutrinos, which is given by the well-known ``see-saw" formula
$M_{\nu} = - N^T M_R^{-1} N$. This means that the two mixing elements
of the electron neutrino, $U_{e 2}$ and $U_{e3}$, get no contribution from
diagonalizing $M_{\nu}$, but come entirely from diagonalizing $L$.
Since $L$ is a known matrix in our model, these two mixing elements
are predicted. In the case of the mixing of the mu and tau neutrinos,
$U_{\mu 3}$ does receive a contribution from diagonalizing 
$M_{\nu}$. However, as can be seen from the form of $N$ this is an effect
of $O(\epsilon)$. The contribution to $U_{\mu 3}$ coming from
diagonalizing $L$, on the other hand, is of order unity, since it
arises from the large parameter $\sigma$. Thus $U_{\mu 3}$ is predicted,
although not precisely.

Since we have written fourteen quantities in terms of five parameters,
there are altogether nine predictions of the model.
Which quantities one takes as ``predicted" depends on which
quantities are used to determine the values of the parameters. We will
use the lepton mass ratios and the angles $V_{cb}$ and $V_{us}$
for this purpose as they are the best measured. As one can see from
the third and eighth of Eqs. (20), one can get the value
$\sigma$ from the ratio $3 V_{cb}^0/(m_{\mu}^0/m_{\tau}^0)$. 
One finds (of course, taking the renormalization effects into
account as was done in \cite{abb}) that numerically $\sigma \simeq 
\sqrt{3}$. Substituting this
into the expression for $m_{\mu}^0/m_{\tau}^0$,  one obtains that
$\epsilon \simeq 0.14$. This is the small parameter of the model that
is responsible for the hierarchy between the second and third families,
and is small enough that the expressions in Eqs. (20) are fairly
accurate. One can use $m_e/m_{\mu}$ and $V_{us}$ to determine
$t_L$ and $t_R$. A careful fit, described later, gives $t_L = 0.236$
and $t_R = 0.205$. That $t_L \simeq t_R$ is easily understood
from Eq. (18) and the well-known Weinberg-Wilczek-Zee-Fritzsch 
result \cite{w-w-z-f} that the Cabbibo angle is  
well acounted for by symmetric mass matrices for the first two families; 
cf. Eq. (1).  The near equality of $t_L$ and $t_R$ is also apparant from the
seventh and ninth relations of Eqs. (20) and the fact that numerically
$V_{us} \cong \sqrt{m_d/m_s}$. The phase factor $e^{i \theta}$ will be
determined from the CP-violating phase $\delta_{CP}$.

The nine predictions, then, are the following. To begin with, there are 
the three famous predictions, {\bf (1)} $m_b^0/m_{\tau}^0 \cong 1$, 
{\bf (2)} $m_s^0 \cong \frac{1}{3} m_{\mu}^0$, and
{\bf (3)} $m_d^0 \cong 3 m_e^0$. The first is the ``good"
prediction of minimal $SU(5)$ unification, and the latter two are the
Georgi-Jarlskog relations. These predictions are manifest from the
first, third, fourth, sixth, and seventh of Eqs. (20).
It is hardly surprising that the model gives these relations, since 
we were guided by them in constructing the model. The fourth prediction
is {\bf (4)} $m_u^0/m_t^0 = 0$. Even if the $u$ quark is not exactly 
massless this relation is a very good approximation to reality. If one 
takes the favored value of $m_u \approx 4$ MeV, then, with reasonable 
assumptions about thresholds in doing the running up to the GUT scale, 
one obtains $\eta \simeq m_u^0/m_t^0 \approx 8 \times 10^{-6}$. This 
is far smaller than any other interfamily ratio of masses. For instance,
the comparable ratio for down quarks is $m_d^0/m_b^0 \cong 10^{-3}$,
and for leptons is $m_e^0/m_{\tau}^0 \cong 3 \times 10^{-4}$. Like the
previous three relations, $m_u \approx 0$ is a reflection of basic
group-theoretical aspects of the model. It comes from the fact, 
explained above, that the $\delta$ and $\delta'$ entries only appear 
in $D$ and $L$.

The remaining five predictions are not simple group-theoretical 
relations like the foregoing, but are non-trivial quantitative
predictions. They are predictions for {\bf (5)} $m_c$, {\bf (6)}
$V_{ub}$, {\bf (7)} $U_{\mu 3}$, {\bf (8)} $U_{e 2}$, and {\bf (9)}
$U_{e3}$. 

The prediction for $m_c$ is particularly interesting. We see immediately
that, for reasons having to do with the group-theoretic structure of
the model, the ratio $m_c^0/m_t^0$ is much less than the corresponding
ratio $m_s^0/m_b^0$ for the down quarks because it is of higher order in 
the small parameter $\epsilon$. This is a highly significant success,
because the minimal Yukawa terms of $SO(10)$ notoriously give these 
ratios to be equal. Moreover, the success is not merely a qualitative
one. When $\epsilon$ and $\sigma$ are fit (using $V_{cb}$ and 
$m_{\mu}/m_{\tau}$) and the renormalization effects are later taken into
account, it is found that $m_c$ comes out within about 5\% of the
experimentally preferred value, which is quite remarkable given
the various experimental and theoretical uncertainties. This success is 
non-trivial, because the reasoning that led to the forms of the mass 
matrices did not depend upon the value of $m_c$, and hence it could have 
been expected that $m_c$ would come out wrong by a large factor. 

Another non-trivial quantitative hurdle for the model is the prediction 
for $V_{ub}$. The eighth, ninth, and tenth relations of Eqs. (20) give
$V_{ub}^0 \cong V_{us}^0 V_{cb}^0 \frac{1}{\sigma^2}(\sqrt{\sigma^2 +
1} \frac{t_R}{t_L} e^{- i \theta} - 1)$. 
If we use the facts that $\sigma \simeq \sqrt{3}$
and $t_L \simeq t_R$, this gives $V_{ub} \simeq V_{us} V_{cb}
(\frac{2}{3} e^{- i \theta} - \frac{1}{3})$. 
A careful fit gives 

\begin{equation}
V_{ub} = V_{us} V_{cb} (0.558 e^{-i \theta} - 0.315).
\end{equation}

\noindent
In other words
the model predicts that $V_{ub}$ should lie on a certain circle
in the complex plane. As can be seen from Fig. 3, the circle for
$V_{ud}V^*_{ub}$ slices neatly through the middle of the presently allowed 
region. Again, this is a very significant success, since the reasoning that 
led to the forms in Eq. (15) was not based on the value of $V_{ub}$.

The prediction for the mixing of $\nu_{\mu}$ and $\nu_{\tau}$ 
has already been discussed. It is one of the key successes of this model 
that this mixing turns out to be nearly maximal. The fact that 
$\sigma \simeq \sqrt{3}$ tells us that the first term in the 
expression for $U_{\mu 3}^0$ in Eq. (20) corresponds to an angle near
$\pi/3$. As we shall see in the next Section, the $O(\epsilon)$
corrections easily bring this down close to the maximal mixing value of
$\pi/4$.

The prediction of this model for the mixing of $\nu_e$ and 
$\nu_{\mu}$ with $\eta = 0$ is quite interesting. From the sixth
relation of Eqs. (20) and the fact that $t_L \simeq t_R$, one sees that
$\frac{1}{3} t_R \simeq \sqrt{m_e/m_{\mu}}$. Thus the model predicts 
that $U_{e 2} \cong \cos \theta_{\mu \tau} \sqrt{m_e/m_{\mu}}$. 
The factor of $\cos \theta_{\mu \tau}$ is crucial \cite{bando}
since without it one would
have $\sin^2 2 \theta_{\rm solar} = 4|U_{e1}|^2|U_{e2}|^2 \approx 4 
(m_e/m_{\mu}) \cong 2 \times 10^{-2}$, 
which is about twice the value needed for the small-angle MSW solution 
to the solar neutrino problem. Since atmospheric neutrino data
tells us that $\cos \theta_{\mu \tau} \simeq 1/\sqrt{2}$, the
model gives just the correct value for the small-angle MSW
solution.

In the future both $V_{ub}$ and $\sin^2 2 \theta_{\rm solar}$ will be 
known better and will provide a sharp test of the model.  The
theoretical uncertainties in the predictions for $V_{ub}$ and $U_{e2}$
are estimated to be only a few percent.

In discussing the $\nu_e - \nu_{\mu}$ mixing above, we have assumed 
that $\eta = 0$. If $\eta$ does not vanish, but is around $8 \times 
10^{-6}$, corresponding to $m_u \approx 4.5$ MeV, then it turns out that 
both the small-angle MSW solution we just discussed and bimaximal 
mixing are possible. This will be discussed in detail in the next 
Section.

Finally, there is the prediction of $\nu_e -\nu_{\tau}$ mixing.
One sees from Eq. (20) that there is a prediction that $U_{e3}
\cong \tan \theta_{\mu \tau} U_{e2} \cong 0.05$.  It is interesting
that even for the bimaximal mixing case that will be discussed
in the next Section, the numerical value of $U_{e3}$ is
virtually unaffected. Thus this prediction is a ``robust" one of
this model.

\section{NEUTRINO MIXING}

\subsection{Mixing of $\nu_{\mu}-\nu_{\tau}$}

In analyzing the predictions of this model for $\nu_{\mu}-\nu_{\tau}$ 
mixing, we may make the approximation that $\eta =0$. This means 
that $N$ has vanishing first row and column. Therefore, in computing
$M_{\nu} = - N^T M_R^{-1} N$ the first row and column of $M_R^{-1}$
are irrelevant. Thus we may write $M_R^{-1}$ as

\begin{equation}
M_R^{-1} = \left( \begin{array}{ccc}
- & - & - \\ - & X & Y \\ - & Y & Z \end{array}
\right),
\end{equation}

\noindent 
where $X$, $Y$, and $Z$ are in general complex. There are consequently
five real parameters (the over all phase does not matter) that come
into the masses and mixing of $\nu_{\mu}$ and $\nu_{\tau}$ from $M_R$. 
As observed earlier, this does not prevent us from making a
qualitative prediction for the mixing parameter $U_{\mu 3}$, since
the contribution to it from diagonalizing the mass matrix 
$M_{\nu}$ is only of order $\epsilon$ and $U_{\mu 3}$ comes 
predominantly from diagonalizing the known matrix $L$. However,
in order to see if more precise predictions can be obtained, we 
shall look at two simple special cases:

\begin{equation} (I) \;\;\;\;\;\; M_R = \left( \begin{array}{ccc} - & 0 & 0 \\ 0
& 0 & B e^{i \beta} \epsilon \\ 0 & B e^{i \beta} \epsilon & 1 \end{array}
\right) \Lambda_R e^{i \gamma}, \end{equation}

\noindent
and

\begin{equation}
(II) \;\;\;\;\;\; M_R = \left( \begin{array}{ccc}
- & 0 & 0 \\ 0 & B e^{i \beta} \epsilon^2 & 0 \\
0 & 0 & 1 \end{array} \right) \Lambda_R e^{i \gamma}.
\end{equation}

\noindent
In these cases only three parameters in $M_R$,
namely $\Lambda_R$, $B$ and $e^{i \beta}$, contribute to the neutrino
observables of the second and third families, since $\epsilon$ has appeared
previously and is used as a natural scaling parameter. In the first case, 

\begin{equation} 
M^I_{\nu} = - \left( \begin{array}{ccc}
0 & 0 & 0 \\ 0 & 0 & \epsilon \\ 0 & \epsilon & 2 + B^{-1} 
e^{- i \beta} \end{array} \right) \frac{M_U^2}{B \Lambda_R} 
e^{-i (\beta + \gamma)}.
\end{equation}

\noindent
The neutrino mixing matrix $U$, now known as the MNS mixing matrix \cite{MNS}, 
is given by
$U = U_L^{\dag} U_{\nu}$,
where $U_L$ is the unitary matrix that diagonalizes $L^{\dag} L$, and
$U_{\nu}$ is the unitary matrix that diagonalizes 
$M_{\nu}^{\dag} M_{\nu}$. For case I, $U_{\nu}$ is given by

\begin{equation}
U_{\nu} = \left( \begin{array}{ccc}
1 & 0 & 0 \\ 0 & \cos \theta_{23}^{\nu *} & \sin \theta_{23}^{\nu} 
\\ 0 & - \sin \theta_{23}^{\nu *} & \cos \theta_{23}^{\nu} \end{array} 
\right),
\end{equation}

\noindent
where  $\tan 2 \theta_{23}^{\nu} = 2 \epsilon/K$, and $K \equiv
2 + B^{-1} e^{i \beta}$. The ratio of eigenvalues of $M_{\nu}$
gives $m_{\nu_2}/m_{\nu_3} \cong 
(\epsilon^2/\left| K \right|^2) (1 - \epsilon^2/\left| K \right|^2 
+ ...)$. One can choose $\cos \theta_{23}^{\nu}$ to be real, and one can write

\begin{equation}
\sin \theta_{23}^{\nu} \cong \sqrt{m_{\nu_2}/m_{\nu_3}} 
e^{-i \xi}\left[1 + 
(\frac{1}{2} - e^{-2 i \xi}) \frac{m_{\nu_2}}{/m_{\nu_3}}\right].
\end{equation}

\noindent
where $e^{i \xi}$ is the phase of $K$.
One readily sees from the form of the charged-lepton mass matrix $L$ in 
Eq. (15) that $\sin \theta_{23}^L \equiv (U_L)_{23}$ is given by
$\tan 2 \theta_{23}^L = - \frac{2 ( \sigma + \epsilon)}{\sigma^2 - 1
+ 2 \sigma \epsilon} = -\frac{2 \sigma}{\sigma^2 -1} + O(\epsilon)$.
Since $\sigma \simeq \sqrt{3}$ it is evident that $\theta_{23}^L \simeq
60^{\circ}$. Using the best fit values of $\sigma$ and $\epsilon$ one
finds, more precisely, that $\theta_{23}^L \cong 63^{\circ}$.

Altogether, then, the mixing parameter of $\nu_{\mu}$ and $\nu_{\tau}$
is given by

\begin{equation}
\begin{array}{lcl}
U_{\mu 3} & \equiv & \sin \theta_{\mu \tau} \\
& & \\
& = & - \sin \theta_{23}^L \cos \theta_{23}^{\nu} + \cos 
\theta_{23}^L 
\sin \theta_{23}^{\nu} \\ & & \\
& \cong & - 0.898 (1 - m_{\nu_3}/m_{\nu_2})
+ 0.441 \sqrt{m_{\nu_3}/m_{\nu_2}} e^{-i \xi}
\end{array}
\end{equation}

\noindent
If neutrino masses are hierarchical, and atmospheric neutrino 
oscillations are $\nu_{\mu}-\nu_{\tau}$ oscillations, then 
$m_{\nu_3} \simeq 0.06$ eV. And if one further assumes the 
small-angle MSW solution
to the solar neutrino problem, then $m_{\nu_2} \simeq 0.003$ eV.
Thus, $m_{\nu_2}/m_{\nu_3} \simeq 0.05$, within a factor of two
or so. Taking it to have the value 0.05, and the phase $\xi$ to vanish,
Eq. (28) gives $U_{\mu 3} \cong -0.756$, and $\sin^2 2 \theta_{\mu \tau}
\cong 0.984$. With the same value of the neutrino mass ratio and 
$\xi$ taken to be $\pi/4$, $\sin^2 2 \theta_{\mu \tau} \cong 0.943$.
We see that there is excellent agreement with the experimental
limits from SuperKamiokande if the complex phase is not too large.
But if $\xi = \pi/2$, with the same mass ratio, $\sin^2 2
\theta_{\mu \tau} \cong 0.77$.

The value of $m_{\nu_2}/m_{\nu_3} = 0.05$ corresponds to
$|K| = 0.63$. Since $B^{-1} e^{i \beta} = |K| e^{i \xi}
-2$, for $\xi = 0$ this gives $B = 0.73$, or $B \epsilon = 0.1$.
In other words, no very great hierarchy is required in $M_R$.

Turning now to case II, we have that 

\begin{equation} 
M^{II}_{\nu} = - \left( \begin{array}{ccc}
0 & 0 & 0 \\ 0 & \epsilon^2 & \epsilon \\ 
0 & \epsilon & 1 + B^{-1} 
e^{- i \beta} \end{array} \right) \frac{M_U^2}{\Lambda_R} 
e^{-i \gamma}.
\end{equation}

\noindent
Consequently, for this case

\begin{equation}
\tan 2 \theta_{23}^{\nu} \cong 2 \epsilon/K',
\end{equation}

\noindent
where $K' \equiv
1 + B^{-1} e^{i \beta}$. The ratio of eigenvalues of $M_{\nu}$
gives $m_{\nu_2}/m_{\nu_3} \cong 
\epsilon^2 \sqrt{1 - |K' -1|^2}/|K'|^2$. If we take,
$m_{\nu_2}/m_{\nu_3} = 0.05$, as before, and assume that $K'$
is real, we have that $K' \cong 0.6$. This gives $\theta_{23}^{\nu} \cong
12.3^{\circ}$, $\theta_{\mu \tau} \cong 50.5^{\circ}$,
and $\sin^2 2 \theta_{\mu \tau} \cong 0.96$. Again, there is good
agreement with the SuperKamiokande results. Moreover, since this
value of $K'$ corresponds to $B = -2.5$, or $B \epsilon^2 \cong
-0.05$, we see that in this case also no great hierarchy is needed in
$M_R$. 

The foregoing discussion is all based on the assumption that the
mixing with the first family is small, so that one has the small-angle
MSW solution to the solar neutrino problem. This will certainly
be the case if $\eta = 0$. As we will now see, if instead $\eta \cong 8
\times 10^{-6}$, as needed to have $m_u \cong 4.5$ MeV, either
the small mixing of $\nu_e$ that we have been considering or large
mixing of $\nu_e$ is possible, depending on the form of $M_R$.

\subsection{Mixing of the First Family}

In the previous discussion, we set $\eta = 0$ in which case,
no matter what the form of $M_R$, the matrix $M_{\nu} = 
- N^T M_R^{-1} N$ has vanishing first row and column, and  
the matrix $U_{\nu}$ that diagonalizes $M_{\nu}^{\dag} M_{\nu}$ has the 
form of Eq. (26).  It is easy to show that the matrix $U_L$ which 
diagonalizes $L^{\dag} L$ has the form

\begin{equation}
U_L \cong \left( \begin{array}{ccc}
\cos \theta_{12}^L & - \sin \theta_{12}^L & 0 \\
\sin \theta_{12}^L & \cos \theta_{12}^L & 0 \\ 
0 & 0 & 1 \end{array} \right) \left( \begin{array}{ccc}
1 & 0 & 0 \\ 0 & \cos \theta_{23}^L & - \sin \theta_{23}^L \\
0 &  \sin \theta_{23}^L & \cos \theta_{23}^L \end{array}
\right),
\end{equation}

\noindent
where $\sin \theta_{12}^L \cong \frac{1}{3} t_R$, $t_R$ is
defined in Eq. (17), and $\theta_{23}^L$ is given after Eq. (27).
Putting these together, one has that the total mixing matrix of
the neutrinos, $U = U_L^{\dag} U_{\nu}$, is

\begin{equation}
U =  \left( \begin{array}{ccc} \cos \theta_{12}^L &
- \sin \theta_{12}^L \cos \theta_{\mu \tau} & - \sin \theta_{12}^L
\sin \theta_{\mu \tau} \\ \sin \theta_{12}^L & \cos \theta_{12}^L 
\cos \theta_{\mu \tau} & \cos \theta_{12}^L \sin \theta_{\mu \tau} \\
0 & - \sin \theta_{\mu \tau} & \cos \theta_{\mu \tau} \end{array}
\right),
\end{equation}

\noindent
where $\theta_{\mu \tau} = \theta_{23}^L - \theta_{23}^{\nu}$.
This yields the results, already given in Eq. (20), for $U_{e2}$ and
$U_{e3}$.

Now we will consider what happens under
what is presumably the more realistic assumption that $\eta \cong
8 \times 10^{-6}$. 

With $\eta \neq 0$, there are two basic possibilities to consider. 
One possibility is that $M_R$ has a form in which its 12, 21, 13, 
and 31 elements all vanish or are negligibly small. If such is the 
case, then the previous analysis applies, and the mixing of
$\nu_e$ is due entirely to the matrix $L$. The only effect of the
parameter $\eta$ in the lepton sector is then to give $\nu_1$ a
mass of about $4 \times 10^{-7}$ eV. The second possibility is that
$M_R$ does have significant 12, 21 and/or 13, 31 elements. If this 
is the case then a strikingly different situation can arise \cite{ab5}, 
namely ``bimaximal" mixing \cite{bimax}, \cite{bimaximal}.

We will first illustrate what happens with a simple example.
Consider the following form for $M_R$:

\begin{equation}
M_R = \left( \begin{array}{ccc}
0 & A \epsilon^3 & 0 \\
A \epsilon^3 & B \epsilon^2 & 0 \\
0 & 0 & 1 \end{array} \right) \Lambda_R.
\end{equation}

\noindent
We normalize $A$ and $B$ by powers of $\epsilon$ simply for later
convenience. The mass matrix of light neutrinos resulting from this form is  

\begin{equation}
M_{\nu} = - N^T M_R^{-1} N = \left(
\begin{array}{ccc} \frac{\eta^2}{\epsilon^4} \frac{B}{A^2} & 0 &
- \frac{\eta}{\epsilon^2} \frac{1}{A} \\
0 & \epsilon^2 & \epsilon \\
- \frac{\eta}{\epsilon^2} \frac{1}{A} & \epsilon & 1  
\end{array} \right) \frac{M_U^2}{\Lambda_R}.
\end{equation}

\noindent
One sees that the 2-3 block has vanishing determinant, so that a rotation
in the 2-3 plane by an angle $\theta_{23}^{\nu} \cong \epsilon$ brings
$M_{\nu}$ to the form

\begin{equation}
M'_{\nu} \cong \left( \begin{array}{ccc}
\frac{\eta^2}{\epsilon^4} \frac{B}{A^2} & \frac{\eta}{\epsilon}
\frac{1}{A} & - \frac{\eta}{\epsilon^2} \frac{1}{A} \\
\frac{\eta}{\epsilon} \frac{1}{A} & 0 & 0 \\
- \frac{\eta}{\epsilon^2} \frac{1}{A} & 0 & 1
\end{array} \right) \frac{M_U^2}{\Lambda_R}.
\end{equation}

\noindent
This can be put in a more transparent form by a rotation in the 
1-3 plane by an angle $\theta_{13}^{\nu} \cong - 
\eta/(\epsilon^2 A)$.
This angle is less than or of order $10^{-4}$ and thus negligible,
practically speaking, so

\begin{equation}
M^{\prime \prime}_{\nu} \cong - \left(
\begin{array}{ccc}
\frac{\eta^2}{\epsilon^4}\frac{(B-1)}{A^2} & \frac{\eta}{\epsilon} 
\frac{1}{A} & 0 \\ \frac{\eta}{\epsilon} \frac{1}{A} & 0 & 0 \\
0 & 0 & 1 \end{array} \right) \frac{M_U^2}{\Lambda_R}.
\end{equation}

\noindent
It is clear that the 11 element, being higher order in $\eta$, is
likely to be much smaller than the 12 and 21 elements. The condition
for this to be the case is that $A/(B-1) > \eta/\epsilon^3 \cong 2 
\times 10^{-3}$. If this very weak condition is satisfied, then
the form of the matrix manifestly corresponds to the situation in which
the $\nu_e$ and $\nu_{\mu}$ together form a pseudo-Dirac pair.
That in turn would mean that the mixing of these two neutrinos
is very close to maximal. 

One sees from Eq. (36) that $m_{\nu_3} = M_U^2/\Lambda_R$, 
and that the splitting between $m_{\nu_1}$ and $m_{\nu_2}$ is
given by $\Delta m^2_{21} \cong 2(\eta^3/\epsilon^5) ((B-1)/A^3) (M_U^2/
\Lambda_R)^2$. For the vacuum solution to the solar neutrino problem, one has
$\Delta m^2_{21} \simeq 10^{-10}$ eV$^2$, so that
$\Delta m^2_{21}/m^2_{\nu_3} \simeq 3 \times 10^{-8}
\cong 2 (\eta^3/\epsilon^5) (B-1)/A^3$. This gives
$A(B-1)^{1/3} \approx 0.06$. Thus no great hierarchy is required
in $M_R$ to get the vacuum oscillation solution. The reason for this 
is that in this scheme the smallness of $\Delta m^2_{21}$ is
due to the extreme smallness of the parameter $\eta$, which
is equal to the ratio $m_u/m_t$. 

It is easy to see from what has already been said that
the matrix $U_{\nu}$ needed to diagonalize $M_{\nu}^{\dag} M_{\nu}$ 
is of the form

\begin{equation}
U_{\nu} \cong \left( \begin{array}{ccc}
1 & 0 & 0 \\ 0 & \cos \theta_{23}^{\nu} & \sin \theta_{23}^{\nu} \\
0 & - \sin \theta_{23}^{\nu} & \cos \theta_{23}^{\nu} \end{array} \right)
\left( \begin{array}{ccc} 1/\sqrt{2} & 1/\sqrt{2} & 0 \\
- 1/\sqrt{2} & 1/\sqrt{2} & 0 \\ 0 & 0 & 1 \end{array} \right).
\end{equation}

\noindent
where we have neglected the tiny rotation $\theta_{13}^{\nu}$.
The matrix $U_L$ is already given in Eq. (31), so that the full
neutrino mixing matrix can be written

\begin{equation}
U \cong U_{SMA} \cdot \left( \begin{array}{ccc} 1/\sqrt{2} & 
1/\sqrt{2} & 0 \\ -1/\sqrt{2} & \sqrt{2} & 0 \\ 0 & 0 & 1 \end{array}
\right),
\end{equation}

\noindent
where $U_{SMA}$ is given in Eq. (32), and is just the form that
results in the small mixing angle (SMA) MSW case of this model. In other
words, the net result of the large mxing of the first family
produced by the $A$ entries in Eq. (34) is simply to multiply
the (SMA) MSW form of $U$ on the right by a rotation of
$\pi/4$ in the 1-2 plane. Consequently, the predictions for $U_{\mu 3}$
and $U_{e3}$ are essentially unaffected. However, $U_{e2}$ becomes
$1/\sqrt{2}$ instead of the value given in Eq. (20).

The interesting lesson is that ``bimaximal mixing" is easy to achieve
if the large mixing of $\nu_{\mu}$ and $\nu_{\tau}$ comes from the
charged lepton sector, i.e., from diagonalizing $L$, while
the large mixing of $\nu_e$ comes from diagonalizing $M_{\nu}$.

The simple form given in Eq. (33) gives $\theta^{\nu}_{23} \cong
\epsilon \cong 8^{\circ}$, and thus $\theta_{\mu \tau} =
\theta^L_{23} - \theta^{\nu}_{23} \cong 55^{\circ}$, corresponding
to $\sin^2 2 \theta_{\mu \tau} = 0.88$. Somewhat larger values
of $\sin^2 2 \theta_{\mu \tau}$ can arise from a more general form
$M_R$.
Consider, for example, 

\begin{equation}
M_R = \left( \begin{array}{ccc} 0 & A \epsilon^3 & C \epsilon^2 \\
A \epsilon^3 & B \epsilon^2 & 0 \\ C \epsilon^2 & 0 & 1 \end{array}
\right) \Lambda_R.
\end{equation}

\noindent
Then 

\begin{equation}
M_{\nu} = - \left( \begin{array}{ccc}
\frac{\eta^2}{\epsilon^4} B & \frac{\eta}{\epsilon} BC &
\frac{\eta}{\epsilon} (BC-A) \\ \frac{\eta}{\epsilon} BC &
\epsilon^2 A^2 & \epsilon A(A+C) \\ \frac{\eta}{\epsilon} (BC-A) &
\epsilon A(A+C) & (A+C)^2 \end{array} \right) (A^2 + B C^2)^{-1}
\frac{M_U^2}{\Lambda_R
}
\end{equation}

\noindent
Here, as in Eq. (34), the 2-3 block has vanishing determinant. The
crucial difference is that the diagonalization of this matrix
involves a rotation in the 2-3 plane by an angle $\theta_{23}^{\nu}
\cong \epsilon \frac{A}{A+C}$. With $C = - \frac{1}{2} A$, for instance,
$\theta_{\mu \tau}$ comes out very close to $45^{\circ}$.
Otherwise, this case is quite similar to that of Eq. (33). 

\section{DETAILS OF A SPECIFIC MODEL}

In the previous Sections we have presented the construction of our $SO(10)$
minimal Higgs model in the framework of effective $SO(10)$ and $SU(5)$
operators.  We now show that one can construct a technically-natural realization 
of this scheme by introducing sets of Higgs and matter superfields with 
a well-defined family symmetry.  We first address the Higgs sector.

\subsection{Higgs Sector with $U(1) \times Z_2 \times Z_2$ Family Symmetry}

The doublet-triplet splitting problem in $SU(5)$, and therefore $SO(10)$,
arises because the colored Higgs in the ${\bf 5} - \overline{\bf 5}$ pairs
of each ${\bf 10}_H$ must be made superheavy at the GUT scale, while just
one pair of Higgs doublets should remain massless there
and be free to develop VEV's at the electroweak scale.  This problem has
been addressed and solved in \cite{b-r} by the introduction of just one
${\bf 45}$ adjoint Higgs field with its VEV pointing in the $B - L$ direction,
together with two pairs of ${\bf 16} + \overline{\bf 16}$ spinor Higgs
fields, two ${\bf 10}$ Higgs in the vector representation plus several 
Higgs singlets.  We shall briefly summarize the solution, but first we 
note that it is necessary to introduce several more Higgs fields in 
the vector and singlet representations in order to generate the Yukawa
structure for the fermion mass matrices presented earlier.  

The authors of \cite{b-r} found that the Higgs superpotential required to 
solve the doublet-triplet splitting problem could be neatly obtained from
their list of Higgs fields by introducing a global family symmetry
group of the type $U(1) \times Z_2 \times Z_2$, which can arise in a 
natural fashion from string theory.  With this in mind, we now list in 
Table I all the Higgs fields to be considered together with their family 
charge assignments.

$$\begin{tabular}{ll}
\multicolumn{2}{l}{\bf Higgs\ Fields\ Needed\ to\ Solve\ 
        the\ 2-3\ Problem:}\\[0.1in]
        ${\bf 45}_{B-L}$: & $A(0)^{+-}$ \\
        ${\bf 16}$: & $C(\frac{3}{2})^{-+},\ 
                C'(\frac{3}{2}-p)^{++}$\\
        $\overline{\bf 16}$: & $\bar{C}(-\frac{3}{2})^{++},
                \ \bar{C}'(-\frac{3}{2}-p)^{-+}$\\
        ${\bf 10}$: & $T_1(1)^{++},\ T_2(-1)^{+-}$\\
        ${\bf 1}$: & $X(0)^{++},\ P(p)^{+-},\ Z_1(p)^{++},
                \ Z_2(p)^{++}$\\[0.1in]
\multicolumn{2}{l}{\bf Additional Higgs\ Fields\ for\ the\ Mass\ Matrices:}
        \\[0.1in]
        ${\bf 10}$: & $T_0(1+p)^{+-},\ T'_o(1+2p)^{+-}$,\\
                & \ $\bar{T}_o(-3+p)^{-+}, \bar{T}'_o(-1-3p)^{-+}$\\
        ${\bf 1}$: & $Y(2)^{-+},\ Y'(2)^{++},\ 
                S(2-2p)^{--},\ S'(2-3p)^{--}$,\\
                & $V_M(4+2p)^{++}$\\[0.1in]
\multicolumn{2}{c}{Table I.  Higgs superfields in the proposed model.}
\end{tabular}$$

\noindent
As noted in the table, in order to complete the construction of the Dirac
mass matrices, four more vector Higgs fields and four additional Higgs 
singlets are needed, while one Higgs singlet is introduced to generate
the right-handed Majorana neutrino mass matrix.

It is then possible to write down explicitly the full Higgs superpotential
from the Higgs $SO(10)$ and family assignments, where we have written
it as the sum of five terms:

\begin{equation}
$$\begin{array}{rcl} 
        W_{\rm Higgs} &=& W_A + W_{CA} + W_{2/3} + W_{H_D} + W_R\\[0.1in]
        W_A &=& tr A^4/M + M_A tr A^2\\[0.1in]
        W_{CA} &=& X(\overline{C}C)^2/M^2_C + F(X) \\
         & & + \overline{C}'(PA/M_1 + Z_1)C + \overline{C}(PA/M_2 + 
          Z_2)C'\\[0.1in]
        W_{2/3} &=& T_1 A T_2 + Y' T^2_2\\[0.1in]
        W_{H_D} &=& T_1 \overline{C}\overline{C} Y'/M + 
          \overline{T}_0 C C' + \overline{T}_0(T_0 S + T'_0 S')\\[0.1in]
        W_R &=& \overline{T}_0 \overline{T}'_0 V_M\\
\end{array}$$
\end{equation}

\noindent
The Higgs singlets are all assumed to develop VEV's at the GUT scale.
We can then determine the fate of the other Higgs fields from the F-flat 
and D-flat conditions.  In particular, $W_A$ fixes $\langle A \rangle$ through
the $F_A = 0$ condition where one solution is $\langle A \rangle \propto B-L$,
the Dimopoulos-Wilczek solution \cite{dim-wil}.  $W_{CA}$ gives a GUT-scale 
VEV to $\overline{C}$ and $C$ by 
the $F_X = 0$ condition and also couples the adjoint $A$ to the spinors 
$C,\ \overline{C},\ C'$ and $\overline{C}'$ without destabilizing the 
Dimopoulos-Wilczek solution or giving Goldstone modes,
as shown in \cite{b-r}.  $W_{2/3}$ gives the 
doublet-triplet splitting by the Dimopoulos-Wilczek mechanism \cite{dim-wil},
\cite{b-r}. $W_{H_D}$
mixes the $(1,2,-1/2)$ doublet in $T_1$ with those in $C'$ 
(by $F_{\overline{C}} = 0$), and in $T_0$ and $T'_0$ (by $F_{\overline{T}_0} 
= 0$).

\subsection{Yukawa Sector}

We now turn to the Yukawa sector and specify the matter fields and their
$U(1) \times Z_2 \times Z_2$ charge assignments which will complete the 
realization of the specific model in question.  For this purpose, we 
require three spinor fields ${\bf 16}_i$, one for each light family,
two vector-like pairs of ${\bf 16} - \overline{\bf 16}$ spinors which 
can get superheavy, a pair of superheavy ${\bf 10}$ fields in the vector 
representation, and three pairs of superheavy ${\bf 1} - {\bf 1}^c$ fermion 
singlets.  The complete listing is given in Table II.

$$\begin{tabular}{lll}
        ${\bf 16}_1(-\frac{1}{2}-2p)^{+-}$ \  & 
                ${\bf 16}_2(-\frac{1}{2}+p)^{++}$ \ & 
                ${\bf 16}_3(-\frac{1}{2})^{++}$ \\
        ${\bf 16}(-\frac{1}{2}-p)^{-+}$ \ & 
                ${\bf 16}'(-\frac{1}{2})^{-+}$ \\
        $\overline{\bf 16}(\frac{1}{2})^{+-}$ \ &
        $\overline{\bf 16}'(-\frac{3}{2}+2p)^{+-}$\\[0.1in]
        ${\bf 10}_1(-1-p)^{-+}$ \ & ${\bf 10}_2(-1+p)^{++}$
          \\[0.1in]
        ${\bf 1}_1(2+2p)^{+-}$ \ & 
                ${\bf 1}_2(2-p)^{++}$ \  
                & ${\bf 1}_3(2)^{++}$\\[0.1in]
        ${\bf 1}^c_1(-2-2p)^{+-}$ \ & ${\bf 1}^c_2(-2)^{+-}$
                \ & ${\bf 1}^c_3(-2-p)^{++}$ \\[0.1in]
\multicolumn{3}{c}{Table II.  Matter superfields in the proposed model.}
  \end{tabular}$$

In terms of these fermion fields and the Higgs fields previously introduced,
one can then spell out all the terms in the Yukawa superpotential which 
follow from their $SO(10)$ and $U(1) \times Z_2 \times Z_2$ assignments:

\begin{equation}
$$\begin{array}{rl}
        W_{Yukawa} &= {\bf 16}_3 \cdot {\bf 16}_3 \cdot T_1 + {\bf 16}_2
                \cdot {\bf 16} \cdot T_1
                + {\bf 16}' \cdot {\bf 16}' \cdot T_1\\
                &+ {\bf 16}_3 \cdot {\bf 16}_1 \cdot T'_0
                + {\bf 16}_2 \cdot {\bf 16}_1 \cdot T_0
                + {\bf 16}_3 \cdot {\overline{\bf 16}} \cdot A\\
                &+ {\bf 16}_1 \cdot \overline{\bf 16}' \cdot Y'
                + {\bf 16} \cdot {\overline{\bf 16}} \cdot P
                + {\bf 16}' \cdot {\overline{\bf 16}}' \cdot S\\
                &+ {\bf 16}_3 \cdot {\bf 10}_2 \cdot C'
                + {\bf 16}_2 \cdot {\bf 10}_1 \cdot C
                + {\bf 10}_1 \cdot {\bf 10}_2 \cdot Y\\
                &+ {\bf 16}_3 \cdot {\bf 1}_3 \cdot \overline{C}
                + {\bf 16}_2 \cdot {\bf 1}_2 \cdot \overline{C}
                + {\bf 16}_1 \cdot {\bf 1}_1 \cdot \overline{C}\\
                &+ {\bf 1}_3 \cdot {\bf 1}^c_3 \cdot Z
                + {\bf 1}_2 \cdot {\bf 1}^c_2 \cdot P
                + {\bf 1}_1 \cdot {\bf 1}^c_1 \cdot X\\
                &+ {\bf 1}^c_3 \cdot {\bf 1}^c_3 \cdot V_M
                + {\bf 1}^c_1 \cdot {\bf 1}^c_2 \cdot V_M\\
  \end{array}$$
\end{equation}

\noindent 
where the coupling parameters have been suppressed.  To obtain the GUT
scale structure for the fermion mass matrix elements, all but the three
chiral spinor fields in the first line of Table II.  will be integrated out 
to yield Froggatt-Nielsen diagrams \cite{f-n} of the type pictured earlier.  
Note that the right-handed Majorana matrix elements will all be generated 
through the Majorana couplings of the $V_M$ Higgs field with conjugate 
singlet fermions given in the last two terms of Eq. (42).

In order to present a clearer description of how the GUT scale mass 
matrices are determined from the Yukawa and Higgs superpotentials, we 
shall illustrate the procedure for the up quark mass matrix $U$.
The three massless color-triplet quark states each with charge $2/3$ are 
obtained as
linear combinations of all such color and charge states within the fermion
supermultiplets given in (41).  In particular, the basis for the left-handed
states $u_L$ and left-handed conjugate states $u^c_L$ can be ordered as 
follows:

\begin{equation}
$$\begin{array}{rl}
  {\cal B}_{u_L} = &\left\{ |3,2,\frac{1}{6}>_{10(16_1)},\
	|3,2,\frac{1}{6}>_{10(16_2)},\ |3,2,\frac{1}{6}>_{10(16_3)},\ 
	|3,2,\frac{1}{6}>_{10(16)}\right.,\\[0.1in]
	& \left.|3,2,\frac{1}{6}>_{10(16')},\ 
	|3,1,\frac{2}{3}>_{\overline{10}(\overline{16})},\ 
	|3,1,\frac{2}{3}>_{\overline{10}(\overline{16}')}\right\}
  \end{array}\\[-0.2in]$$
\end{equation}
\begin{equation}
$$\begin{array}{rl}
  {\cal B}_{u^c_L} = &\left\{ |\bar{3},1,-\frac{2}{3}>_{10(16_1)},\ 
	|\bar{3},1,-\frac{2}{3}>_{10(16_2)},\ |\bar{3},1,-\frac{2}{3}>
		_{10(16_3)},\ 
	|\bar{3},1,-\frac{2}{3}>_{10(16)}\right.,\\[0.1in]
	& \left.|\bar{3},1,-\frac{2}{3}>_{10(16')},\ 
	|\bar{3},2,-\frac{1}{6}>_{\overline{10}(\overline{16})},\ 
	|\bar{3},2,-\frac{1}{6}>_{\overline{10}(\overline{16}')}\right\}
  \end{array}$$
\end{equation}

\noindent
where the states are labeled by their representations and hypercharge 
according to $|SU(3)_c,\ SU(2)_L,\ Y\rangle_{SU(5)(SO(10))}$.

We then form the Yukawa contribution $u^{cT}_L C^{-1}D_u u_L$ by using the 
above bases and the superpotentials to obtain the matrix

$$D_u = \left(\matrix{0 & 0 & 0 & 0 & 0 & 0 & y' \cr
	  0 & 0 & 0 & t_2 & 0 & 0 & 0 \cr
	  0 & 0 & t_3 & 0 & 0 & a & 0 \cr
	  0 & t_2 & 0 & 0 & 0 & p & 0 \cr
	  0 & 0 & 0 & 0 & t' & 0 & s'' \cr
	  0 & 0 & a & p & 0 & 0 & 0 \cr
	  y' & 0 & 0 & 0 & s'' & 0 & 0 \cr}\right)$$

\noindent
where we have introduced the following shorthand notation:

\begin{equation}
$$\begin{array}{rlrl}
  t_3 &= \lambda_{16_3 16_3 T_1}\langle T_1 \rangle, &\quad t_2 =& 
	\lambda_{16_2 16 T_1}\langle T_1 \rangle, 
	\quad t' = \lambda_{16' 16' T_1} \langle T_1 \rangle,\\
  a &= \lambda_{16_3 \overline{16} A}\langle A \rangle, &\quad 
	  p =& \lambda_{16 \overline{16} P}\langle P \rangle,\\
  s'' &= \lambda_{16' \overline{16}' S}\langle S \rangle, &\quad 
	  y' =& \lambda_{16_1 \overline{16}' Y'}\langle Y'
 	  \rangle\\
  \end{array}$$.
\end{equation}

\noindent 
We can then determine from this matrix the three pairs of zero-mass eigenstates
at the GUT scale where the electroweak VEV of $T_1$ vanishes:

\begin{equation}
$$\begin{array}{rl}
  |u_{1L}\rangle =& \left[ |10(16_1)\rangle - \frac{y'}{s''}|10(16')
	\rangle\right]/\sqrt{1 + y'^2/s''^2}\\
  |u_{2L}\rangle =& |10(16_2)\rangle \\
  |u_{3L}\rangle =& \left[ |10(16_3)\rangle - \frac{a}{p}|10(16)\rangle\right]
	/\sqrt{1 + a^2/p^2}\\
  |u^c_{1L}\rangle =& \left[ |10(16_1)\rangle - \frac{y'}{s''}|10(16')
	\rangle\right]/\sqrt{1 + y'^2/s''^2}\\
  |u^c_{2L}\rangle =& |10(16_2)\rangle \\
  |u^c_{3L}\rangle =& \left[ |10(16_3)\rangle - \frac{a}{p}|10(16)\rangle
	\right]/\sqrt{1 + a^2/p^2}\\
    \end{array}$$
\end{equation}

\noindent
and where the states are now simply labeled by their $SU(5)$ and 
$SO(10)$ representations.

Finally, the Dirac matrix $U$ for the three light quark states $u,\ c,\ t$
is obtained by bracketing the electroweak contributions by the appropriate
$u^c_{iL}$ state on the left and the $u_{jL}$ state on the right.  The
result obtained for $U$ has exactly the form found earlier in Eq. (15) 
from the previous effective operator approach, with the 
identifications:

\begin{equation}
$$\begin{array}{rl}
  M_U =& (t_3)_{5(10)}\\
  \epsilon M_U =& |3(a_q/p)(t_2)_{5(10)}|\\
  \eta M_U =& (y'/s'')^2 (t')_{5(10)}\\
  \end{array}$$
\end{equation}

\noindent
Here the subscript on $a_q$ signifies a factor of $1/3$ arising from the 
$B-L$ VEV of the $A$ in the adjoint representation, while the subscripts 
on the $t$ terms specify the appropriate doublet VEV in the ${\bf 10}$ for 
$T_1$.  We have neglected the state normalization factors in (47) but will 
later argue that they can all be taken to be approximately unity.

The Dirac matrices, $D,\ N,\ L$ are constructed in a similar fashion.
In the case of $D$ and $L$, the bases corresponding to Eqs. (43) and (44) are 
enlarged by two states lying in the ${\bf 10}_1$ and ${\bf 10}_2$ 
representations of $SO(10)$.  For $N$, on the other hand, in addition 
to the two above states, one must add the singlet fermion contributions
from the representations ${\bf 1}_k,\ k=1,2,3$ for ${\cal B}_{\nu_L}$
and ${\bf 1}^c_k,\ k=1,2,3$ for ${\cal B}_{\nu^c_L}$.  We list the zero-mass
state vectors for $D$, $N$ and $L$ in analogy to Eqs. (46):

\begin{equation}
$$\begin{array}{rl}
  |d_{1L}\rangle =& \left[ |10(16_1)\rangle - \frac{y'}{s''}|10(16')
	\rangle\right]/\sqrt{1 + y'^2/s''^2}\\
  |d_{2L}\rangle =& |10(16_2)\rangle \\
  |d_{3L}\rangle =& \left[ |10(16_3)\rangle - \frac{a}{p}|10(16)\rangle\right]
	/\sqrt{1 + a^2/p^2}\\
  |d^c_{1L}\rangle =& \left[ |\bar{5}(16_1)\rangle - \frac{y'}{s''}
	|\bar{5}(16')\rangle\right]/\sqrt{1 + y'^2/s''^2}\\
  |d^c_{2L}\rangle =& \left[|\bar{5}(16_2)\rangle - \frac{c}{y}
	|\bar{5}(10_2)\rangle\right]/\sqrt{1 + c^2/y^2}\\
  |d^c_{3L}\rangle =& \left[ |\bar{5}(16_3)\rangle - \frac{a}{p}
	|\bar{5}(16)\rangle\right]/\sqrt{1 + a^2/p^2}\\
    \end{array}$$
\end{equation}

\begin{equation}
$$\begin{array}{rl}
  |n_{1L}\rangle =& \left[ |\bar{5}(16_1)\rangle - \frac{y'}{s''}
	|\bar{5}(16')\rangle\right]/\sqrt{1 + y'^2/s''^2}\\
  |n_{2L}\rangle =& \left[ |\bar{5}(16_2)\rangle - \frac{c}{y}
	|\bar{5}(10_2)\rangle\right]/\sqrt{1 + c^2/y^2}\\
  |n_{3L}\rangle =& \left[ |\bar{5}(16_3)\rangle - \frac{a}{p}
	|\bar{5}(16)\rangle\right]/\sqrt{1 + a^2/p^2}\\
  |n^c_{1L}\rangle =& \left[ |1(16_1)\rangle - \frac{y'}{s''}|1(16')
	\rangle - \frac{\bar{c}_1}{x}|1^c_1\rangle\right]
	/\sqrt{1 + y'^2/s''^2 + \bar{c}^2_1/x^2}\\
  |n^c_{2L}\rangle =& \left[ |1(16_2)\rangle - \frac{\bar{c}_2}{p_{22}}
	|1^c_2\rangle\right]/\sqrt{1 + \bar{c}^2_2/p^2_{22}}\\
  |n^c_{3L}\rangle =& \left[ |1(16_3)\rangle - \frac{a}{p}|1(16)\rangle
	- \frac{\bar{c}_3}{z}|1^c_3\rangle\right]/\sqrt{1 + a^2/p^2 
	+ \bar{c}^2_3/z^2}\\
    \end{array}$$
\end{equation}

\begin{equation}
$$\begin{array}{rl}
  |\ell_{1L}\rangle =& \left[ |\bar{5}(16_1)\rangle - \frac{y'}{s''}
	|\bar{5}(16')\rangle\right]/\sqrt{1 + y'^2/s''^2}\\
  |\ell_{2L}\rangle =& \left[|\bar{5}(16_2)\rangle - \frac{c}{y}
	|\bar{5}(10_2)\rangle\right]/\sqrt{1 + c^2/y^2}\\
  |\ell_{3L}\rangle =& \left[ |\bar{5}(16_3)\rangle - \frac{a}{p}
	|\bar{5}(16)\rangle\right]/\sqrt{1 + a^2/p^2}\\
  |\ell^c_{1L}\rangle =& \left[ |10(16_1)\rangle - \frac{y'}{s''}|10(16')
	\rangle\right]/\sqrt{1 + y'^2/s''^2}\\
  |\ell^c_{2L}\rangle =& |10(16_2)\rangle \\
  |\ell^c_{3L}\rangle =& \left[ |10(16_3)\rangle - \frac{a}{p}
	|10(16)\rangle\right]/\sqrt{1 + a^2/p^2}\\
    \end{array}$$
\end{equation}

\noindent
In the above we have introduced, in analogy with Eqs. (45), the additional
shorthand notation:

\begin{equation}
\begin{array}{rlrl}
  c =& \lambda_{16_210_1C}\langle C \rangle, &\quad \bar{c}_i =& 
	\lambda_{16_i1_i\bar{C}}\langle \bar{C} \rangle, 
	i=1,2,3,\\
  x =& \lambda_{1_1 1^c_1 X}\langle X \rangle, &\quad y =& 
	\lambda_{10_1 10_2 Y}\langle Y \rangle,\\
  z =& \lambda_{1_3 1^c_3 Z}\langle Z \rangle, &\quad p_{22} =& 
	\lambda_{1_2 1^c_2 P}\langle P \rangle\\
  \end{array}$$
\end{equation}

The Dirac matrices $D,\ N$ and $L$ are found by forming matrix elements of the 
electroweak symmetry breaking VEV's with the appropriate basis vectors. 
Again these matrices have exactly the structures given in Eqs. (15),
provided the state normalization factors are approximated by unity, i.e.,
we assume that the zero-mass states have their large components in the 
chiral representations ${\bf 16}_1,\ {\bf 16}_2$ and ${\bf 16}_3$ and that 
all the other components are small.  We shall return to this point in the 
next Section.  In the meantime we make the identifications 

\begin{equation}
$$\begin{array}{rl}
  M_D =& (t_3)_{\bar{5}(10)}\\
  \epsilon M_D =& |3(a_q/p)(t_2)_{\bar{5}(10)}|\\
  \sigma M_D =& -(c/y)(c')_{\bar{5}(16)}\\
  \Delta M_D =& t_0 \bar{t}_0 /s\\
  \delta' e^{i\phi} M_D =& t'_0 \bar{t}_0 /s'\\
  \end{array}$$
\end{equation}

\noindent
in terms of the notation given in Eqs. (45) and (51) and the following:

\begin{equation}
$$\begin{array}{rlrl}
  t_0 =& \lambda_{16_116_2T_0}, &\quad t'_0 =& \lambda_{16_1 16_3T'_0}\\
  \bar{t}_0 =& \lambda_{CC'\bar{T}_0}\langle C \rangle \langle C' \rangle, 
	&\quad c' =& \lambda_{16_310_2C'}\langle C' \rangle,\\
  s =& \lambda_{T_0 \bar{T}_0 S}\langle S \rangle, &\quad s' =& 
	\lambda_{T'_0 \bar{T}_0 S'}\langle S' \rangle\\
  \end{array}$$
\end{equation}

\noindent
The phase $\phi$ appearing in the $\delta'$ term can be understood to 
arise from a phase in the VEV for $S'$.  The structures of the Dirac matrix
elements given in Eqs. (15), (47) and (52) can be understood in terms 
of the simple Froggatt-Nielsen diagrams of Fig. 1 and 2, with the Higgs 
fields labeled as in Table I.

Turning to the right-handed Majorana mass matrix, we use the zero mass 
left-handed conjugate states that were obtained implicitly above for the
Dirac matrix $N$ to form the basis for $M_R$.  
The right-handed Majorana matrix is then obtained by bracketing the 
Majorana Higgs $V_M$ with the appropriate zero mass conjugate neutrino
states in (49).  We obtain 

\begin{equation}
$$ M_R = \left(\matrix{ 0 & A\epsilon^3 & 0 \cr A\epsilon^3 & 0 & 0 \cr 
	0 & 0 & 1 \cr}\right)\Lambda_R$$
\end{equation}

\noindent 
where 

\begin{equation}
$$\begin{array}{rl}
  M_3 =\Lambda_R =& \lambda_{1^c_3 1^c_3 V_M} \langle V_M \rangle 
	(\bar{c}_3/z)^2,\\
  M_2 = - M_1 = A\epsilon^3\Lambda_R =& \lambda_{1^c_1 1^c_2 V_M} 
	\langle V_M \rangle (\bar{c}_1/x)(\bar{c}_2/p_{22})\\
  \end{array}$$
\end{equation}

\noindent 
The lighter two right-handed Majorana masses are degenerate and have 
opposite CP-parity.
Note that the whole right-handed Majorana matrix has been generated 
in this simple model by one Majorana VEV coupling superheavy conjugate
fermion singlets as shown in Fig. 4. 

We conclude this Section with a summary of the GUT scale predictions
derived from the Dirac and Majorana mass matrices with the particular 
parameters appropriate for the model in question.  For convenience we 
give the whole set equations which are the counterpart of Eqs. (19).

\begin{equation}
\begin{array}{rl}
m^0_t/m^0_b \cong & (\sigma^2 + 1)^{-1/2}M_U/M_D,\quad
        m_u^0/m_t^0 \cong \eta, \\
m_c^0/m_t^0 \cong & \frac{1}{9} \epsilon^2 \cdot [1 - \frac{2}{9} 
        \epsilon^2],\quad 
        m_b^0/m_{\tau}^0 \cong 1 - \frac{2}{3} \frac{\sigma}{\sigma^2 + 1} 
        \epsilon, \\
m_s^0/m_b^0 \cong & \frac{1}{3} \epsilon 
        \frac{\sigma}{\sigma^2 + 1}\cdot [1  + \frac{1}{3} \epsilon 
        \frac{1 - \sigma^2 - \sigma \epsilon/3}
        {\sigma (\sigma^2 + 1)} + \frac{1}{2} (t_L^2 + t_R^2) ], \\
m_d^0/m_s^0 \cong & t_L t_R \cdot [ 1 - \frac{1}{3} 
        \epsilon\frac{\sigma^2 + 2}{\sigma (\sigma^2 + 1)} - (t_L^2 + t_R^2)\\
        &\quad + (t_L^4 + t_L^2 t_R^2 + t_R^4)], \\
m_{\mu}^0/m_{\tau}^0 \cong & \epsilon 
        \frac{\sigma}{\sigma^2 + 1} \cdot[1 + \epsilon \frac{1 - 
        \sigma^2 - \sigma \epsilon}{\sigma(\sigma^2 + 1)}
        + \frac{1}{18}(t_L^2 + t_R^2)], \\
m_e^0/m_{\mu}^0 \cong & \frac{1}{9} t_L t_R \cdot 
        [1 - \epsilon \frac{\sigma^2 + 2}{\sigma(\sigma^2 + 1)} + \epsilon^2
        \frac{\sigma^4 + 9 \sigma^2/2 + 3}{\sigma^2(\sigma^2 + 1)^2}\\
        & \quad - \frac{1}{9} (t_L^2 + t_R^2)], \\
V_{cb}^0 \cong & \frac{1}{3} \epsilon 
        \frac{\sigma^2}{\sigma^2 + 1} \cdot [1 + \frac{2}{3} 
        \epsilon \frac{1}{\sigma(\sigma^2 + 1)}], \\
V_{us}^0 \cong & t_L [ 1 -\frac{1}{2} t_L^2 - t_R^2 
        + t_R^4 + \frac{5}{2} t_L^2 t_R^2 + \frac{3}{8} t_L^4\\ 
        & \quad - \frac{\epsilon}{3 \sigma \sqrt{\sigma^2 + 1}} \frac{t_R}{t_L}
        e^{- i \theta}], \\
V_{ub}^0 \cong & \frac{1}{3} t_L \epsilon 
        \frac{1}{\sigma^2 + 1} [\sqrt{\sigma^2 + 1} \frac{t_R}{t_L} 
        e^{-i \theta} (1 - \frac{1}{3} \epsilon \frac{\sigma}{\sigma^2 + 1})\\
        & \quad - (1 - \frac{2}{3}\epsilon \frac{\sigma}{\sigma^2 + 1})], 
        \\[0.1in]
m_2^0/m_3^0 \cong & \left(\frac{\eta}{A\epsilon\sqrt{1 + \epsilon^2}}\right)
        \left[ 1 + \frac{\eta}{A\epsilon^3\sqrt{1 + \epsilon^2}} \right], 
        \\[0.1in]
m_1^0/m_3^0 \cong & \left(\frac{\eta}{A\epsilon\sqrt{1+\epsilon^2}}
        \right)\left[ 1 - \frac{\eta}{2A\epsilon^3\sqrt{1+\epsilon^2}} 
        \right],\\[0.1in]
U_{\mu 3}^0 \cong & - \frac{1}{\sqrt{\sigma^2 + 1}}(\sigma - \epsilon
        \frac{\sigma^2}{\sigma^2 + 1}),\\[0.1in]
U_{e 2}^0 \cong & -\frac{1}{\sqrt{2}}\left[ 1 - \frac{\epsilon}{3\sigma}t_L 
        e^{i\theta}\right.\\
        & \quad + \left. \frac{1}{3\sqrt{\sigma^2 + 1}}(1 + \epsilon\sigma)
        t_R\right],\\[0.1in]
U_{e 3}^0 \cong & \frac{1}{3\sqrt{\sigma^2 + 1}}(\sigma - \epsilon)
        t_R - \frac{\eta}{A\epsilon^2}\\[0.1in]
\end{array}
\end{equation}

\noindent
To the quark equations we have added the ratio $m^0_t/m^0_b$ which 
involves the ratio of $\langle {\bf 5}(T_1) \rangle$ to $\langle {\bf 
\overline{5}}(T_1)\rangle$, i.e., $M_U/M_D$, as well as giving the 
leptonic mass ratios and mixings specific to the model in question.

\subsection{Numerical Evaluation of Matrix Parameters}

We have elaborated above how the simple explicit model proposed gives
precisely the structure for the Dirac mass matrices that was obtained 
from the effective operator approach.  We now show that the entries
are also numerically in the range to fit the quark and lepton mass and mixing
data.  

In order to compare the GUT scale predictions in Eq. (56) with the low  
scale data, the GUT scale values are first evolved from $\Lambda_G = 2 \times 
10^{16}$ GeV down to the SUSY scale which is taken to be $\Lambda_{\rm SUSY}
= m_t(m_t) = 165$ GeV by use of the 2-loop MSSM beta functions.  For this 
purpose, the mass ratios at the two scales are related by the $\eta_{i/j}$ 
running factors, while the quark mixing elements are scaled by the $\eta_{ij}$
factor according to 

\begin{equation}
\begin{array}{rl}
	\left(\frac{m_i}{m_j}\right)_{\rm SUSY} =&\left(\frac{m^0_i}{m^0_j}
		\right)/\eta_{i/j},\\[0.2in]
	(V_{ij})_{\rm SUSY} =&V^0_{ij}/\eta_{ij},\quad (ij) = (ub),\ (cb),
		\ (td),\ (ts)\\
\end{array}
\end{equation}

\noindent
The remaining evolutions to the bottom and charm quark or tau lepton running 
mass scales, or to the 1 GeV scale in the case of the light quarks and 
leptons, is carried out with the 3-loop QCD and 1-loop QED renormalization 
group equations.  Here the running factors are $\eta_i$ with the mass
ratios scaled according to 

\begin{equation}
	m_i(m_i) = (m_i)_{\rm SUSY} /\eta_i(m_t)\\
\end{equation}

\noindent
or similarly, with the running mass scale $m_i$ replaced by 1 GeV.
With $\tan \beta = 5$ used for the numerical evaluations for reasons that 
will become apparently shortly, $\alpha_s(M_Z) = 0.118,\ \alpha(M_Z) =
1/127.9$ and $\sin^2 \theta_W = 0.2315$, the running factors are given by 

\begin{equation}
\begin{array}{rlrl}
  \eta_{u/t} = \eta_{c/t} =&0.6927,\quad & \eta_{d/b} =&\eta_{s/b} = 0.8844\\
  \eta_{\mu/\tau} =&0.9988,  & \eta_{b/t} =& 0.5094\\
  & \multicolumn{3}{l}{\eta_{ub} = \eta_{cb} = \eta_{td} = \eta_{ts} = 0.8853}
	\\[0.1in]
  \eta_u(m_t) = 0.4235,\quad &\eta_c(m_t) = 0.4733,\quad &\eta_t(m_t) = 1.0000\\
  \eta_d(m_t) = 0.4262,\quad &\eta_s(m_t) = 0.4262,\quad &\eta_t(m_t) = 0.6540\\
  \eta_e(m_t) = 0.9816,\quad &\eta_\mu(m_t) = 0.4816,\quad &\eta_\tau(m_t) 
	= 0.9836\\
\end{array}
\end{equation}

\noindent
Finally, finite corrections must be applied to $m_s$, $m_b$ and the 
evolved quark mixing matrix elements which arise from gluino and chargino
loops.  The correction factors are conventionally written as
$(1 + \Delta_s),\ (1 + \Delta_b)$ and $(1 + \Delta_{cb})$ where we have
used 

\begin{equation}
  \Delta_s = -0.20,\quad \Delta_b = -0.15,\quad \Delta_{cb} = -0.05\\
\end{equation}

\noindent
as explained below.

Using the quantities \cite{data} $m_t(m_t) = 165\ {\rm GeV},\ m_{\tau} = 
1.777\ {\rm GeV},\ m_{\mu} = 105.7\ {\rm MeV},\ m_e = 0.511\ {\rm MeV},\ m_u 
= 4.5\ {\rm MeV},\ V_{us} = 0.220,\ V_{cb} = 0.0395$, and $\delta_{CP} = 64^o$
to determine the input parameters, one obtains for them
$M_U \simeq 113\ {\rm GeV},\ M_D \simeq 1\ 
{\rm GeV},\ \sigma = 1.780,\ \epsilon = 0.145,\ t_L = 0.236,\ t_R = 0.205,
\ \theta = 34^o\ ({\rm corresponding\ to\ } \delta =0.0086,\ \delta' = 
0.0079,\ \phi =54^o)$, and $\eta = 8 \times 10^{-6}$.  With these inputs
the remaining quark masses and mixings are obtained, to be compared with the 
experimental values \cite{data} in parentheses:

\begin{equation}
\begin{array}{rll}
        m_c(m_c) =& 1.23\ {\rm GeV}\qquad & (1.27 \pm 0.1\ {\rm GeV})\\
        m_b(m_b) =& 4.25\ {\rm GeV}\qquad & (4.26 \pm 0.11\ {\rm GeV})\\
        m_s({\rm 1\ GeV}) =& 148\ {\rm MeV}\qquad & (175 \pm 50\ {\rm MeV})\\
        m_d({\rm 1\ GeV}) =& 7.9\ {\rm MeV}\qquad & (8.9 \pm 2.6\ {\rm MeV})\\
        |V_{ub}/V_{cb}| =& 0.080 \qquad & (0.090 \pm 0.008)\\
        \end{array}
\end{equation}

\noindent
where the finite SUSY loop corrections for $m_b,\ m_s$ and $V_{cb}$ have been 
rescaled to give $m_b(m_b) \simeq 4.25$ GeV for $\tan \beta = 5.$  Had we 
chosen $\delta_{CP} = 70^o$ as input, on the other hand, we would find 
instead $|V_{ub}/V_{cb}| = 0.085$.  With the numerical values in (61) we 
find for $\bar{\rho},\ \bar{\eta}$ and the $\alpha,\ \beta$ and $\gamma$
angles of the unitarity triangle pictured in Fig. 3

\begin{equation}
	\bar{\rho} = 0.143,\quad \bar{\eta} = 0.305,\quad
	\alpha = 96^o,\quad \beta = 20^o,\quad \gamma = 64^o\\
\end{equation}

\noindent
The upper vertex of the triangle appears to be circled precisely in the 
allowed experimental region.

Additional predictions follow for the neutrino sector.  The effective light
neutrino mass matrix of Eq. (34) or (36) with $B = 0$ leads to bimaximal 
mixing with a large angle solution for atmospheric neutrino oscillations 
\cite{atm} and the ``just-so'' vacuum solution \cite{bimax} involving two 
pseudo-Dirac neutrinos, if we set $\Lambda_R = 2.4 \times 10^{14}$ GeV and
$A = 0.05$.  We then find 

\begin{equation}
\begin{array}{ll}
  \multicolumn{2}{c}{m_3 = 54.3\ {\rm meV},\ m_2 = 59.6\ {\rm \mu eV},
        \ m_1 = 56.5\ {\rm \mu eV}}\\[0.1in]
  \multicolumn{2}{c}{M_3 = 2.4 \times 10^{14}\ {\rm GeV},\ M_2 = M_1 = 
	3.66 \times 10^{10}\ {\rm GeV}}\\[0.1in]
  \multicolumn{2}{c}{U_{e2} = 0.733,\ U_{e3} = 0.047,\ U_{\mu 3} = -0.818,
        \ \delta'_{CP} = -0.2^o}\\[0.1in]
        \Delta m^2_{23} = 3.0 \times 10^{-3}\ {\rm eV^2},\quad &        
                \sin^2 2\theta_{\rm atm} = 4|U_{\mu 3}|^2|U_{\tau 3}|^2 = 0.89
		\\[0.1in]
        \Delta m^2_{12} = 3.6 \times 10^{-10}\ {\rm eV^2},\quad &
                \sin^2 2\theta_{\rm solar} = 4|U_{e1}|^2|U_{e2}|^2 = 0.99
		\\[0.1in]
        \Delta m^2_{13} = 3.0 \times 10^{-3}\ {\rm eV^2},\quad &
             \sin^2 2\theta_{\rm reac} = 4|U_{e3}|^2(1 - |U_{e3}|)^2 = 0.009\\
        \end{array}
\end{equation}

\noindent
The effective scale of the right-handed Majorana
mass contribution occurs two orders of magnitude lower than the SUSY GUT 
scale of $\Lambda_G = 1.2 \times 10^{16}$ GeV.  The effective two-component
reactor mixing angle given above should be observable at a future neutrino 
factory, whereas the present limit from the CHOOZ experiment \cite{CHOOZ} 
is approximately 0.1 for the above $\Delta m^2_{23}$.  In principle, the
parameter $A$ appearing in $M_R$ can also be complex, but we find that in 
no case does the leptonic CP-violating phase, $\delta'_{CP}$ exceed $10^o$
in magnitude.  Hence the model predicts leptonic CP-violation will be 
unobservable.

The vacuum solar solution depends critically on the appearance of the 
parameter $\eta$ in the matrix $N$, corresponding to the non-zero $\eta$ 
entry in $U$ which gives the up quark a mass at the GUT scale.  Should we 
set $\eta = 0$, only the small-angle MSW solution \cite{MSW} would be 
obtained for the solar neutrino oscillations.  The large angle MSW solution 
is disfavored by the larger hierarchy, i.e., very small $A$ value, 
required in $M_R$.  

Finally we address the issue that the state normalization factors were all
replaced by unity in Eqs. (47) and (52) for the various matrix parameters.
This is a good approximation provided the three fermion spinor states
$|16_1\rangle,\ |16_2\rangle,\ |16_3\rangle$ provide the dominant 
contributions to the zero-mass quark and lepton states at the GUT scale.
In particular, the following ratios must be much less than unity:

\begin{equation}
$$(a/p)^2,\ (y'/s'')^2,\ (c/y)^2,\ (\bar{c}_1/x)^2,\ 
	(\bar{c}_2/p_{22})^2,\ (\bar{c}_3/z)^2 \ll 1$$
\end{equation} 

Let us assume for simplicity that the electroweak couplings of
$\langle T_1 \rangle$
in $t_3,\ t_2$ and $t'$ in Eq. (45) and of $\langle C' \rangle$ in
$c'$ of (53) are identical.  Then with $\epsilon = |3(a_q/p)| = |a_\ell/p| 
= 0.14$, we find the condition $(a/p)^2 \simeq 0.02 \ll 1$ holds. 
To obtain an up quark mass $m_u(1 GeV) \simeq 4.5$ MeV, we need
$\eta \cong (y'/s'')^2 \simeq 8 \times 10^{-6}$ at the GUT 
scale, which easily satisfies (64).

Requiring that $(c/y)^2 \ll 1$ and with the result from Eqs. (52) that
\begin{equation}
$$\sigma \simeq \left|\frac{c}{y} {{\langle \bar{5}(C') \rangle}\over{
	\langle \bar{5}(T_1) \rangle}}\right| \simeq 1.8$$
\end{equation}

\noindent
leads us to the results that 
\begin{equation}
\begin{array}{rl}
\tan \gamma \equiv& {{\langle \bar{5}(C') \rangle}\over{\langle 
	\bar{5}(T_1) \rangle}} \gg \sigma \\[0.2in]
 \tan \beta \simeq& \sqrt{\sigma^2 + 1}m^0_t (cos \gamma)/m^0_b 
	\ll m^0_t/m^0_b\\
  \end{array}
\end{equation}

\noindent
in terms of the $T_1 - C'$ mixing angle, $\gamma$, in Eq. (11).
With $c/y \cong 0.1$, for example, $\tan \gamma \simeq 18$ which implies
$\tan \beta \simeq 6$, a very reasonable mid-range value allowed 
by experiment.  For this reason, we have chosen to illustrate the 
numerical results above with $\tan \beta = 5$.

The remaining ratios in Eq. (64) can also be satisfied.  For comparable
$\bar{c}_i$'s, $A\epsilon^3 \sim 1.4 \times 10^{-4}$ obtained from 
Eq. (55) requires that $\langle Z \rangle /\sqrt{\langle X\rangle
\langle P \rangle} \sim 0.01$.  This ratio is consistent with the VEV's
needed in the Higgs superpotential of Eq. (41) in order to solve the 
doublet-triplet splitting problem.

Turning now to the parameters $\delta$ and $\delta'$, we note that 
the near equality of their magnitudes leads to the ratio 
$\delta/|\delta'| \cong s/|s'| \simeq 1$.  Moreover, if we assume
$y \sim y'$, we obtain the estimate  
$\delta \sim cy'/(ys)\tan \gamma \sim 5 \times 10^{-3}$
with the numbers obtained earlier, whereas the actual value needed
is $\delta \simeq 0.008$.

Thus we have found that not only are the desired forms of the Dirac (and 
Majorana) matrices generated by the model of this Section, but that the
numerical values required for the matrix parameters are also quite reasonable.

\section{SUMMARY}

Both the largeness of the atmospheric neutrino
mixing $U_{\mu 3}$ and the smallness of the quark mixing $V_{cb}$ 
can be elegantly accounted for by the idea that the charged lepton mass
matrix $L$ is highly asymmetric or ``lopsided" and that the down-quark
mass matrix $D$ is related to the transpose of $L$ by an $SU(5)$ symmetry.
This idea was discovered independently by several groups
and has since been used in numerous models of fermion masses. Remarkably,
exactly such mass matrices emerged in our work from quite other considerations 
than neutrino masses and mixings, specifically from an attempt to
construct the simplest possible realistic $SO(10)$ model. 

Advances have been made in recent years in simplifying the Higgs structure
of SUSY $SO(10)$ models. If one assumes the minimal set of Higgs
fields that can break $SO(10)$ down to the standard model group,
the possibilities for Yukawa terms for the quarks and leptons become
significantly restricted. It turns out that there is what seems to be
a uniquely simple set of $SO(10)$ Yukawa terms that gives realistic masses 
and mixings. This set consists of only six effective
Yukawa terms (five if $m_u =0$)
which satisfactorily fits all nine masses of the quarks and charged
leptons as well as the four CKM parameters. In addition, large
$\nu_{\mu}-\nu_{\tau}$ mixing emerges automatically. Moreover, in
this uniquely simple model, the simplest possibilities for
the Majorana mass matrix $M_R$ of the right-handed neutrinos lead either
to small angle MSW values for the solar neutrino mixing or to
vacuum oscillation values. In this paper we have studied in detail
the consequences of different forms of $M_R$ for the neutrino mixing
angles and mass ratios.

In the published literature no more predictive and economical
a model for quark and lepton masses than the one discussed here
exists that is also consistent with present knowledge.
It is striking that in this model a single term and a single 
parameter (which we call $\sigma$) accounts for no less than four puzzling
aspects of the light fermion spectrum: the largeness of $U_{\mu 3}$,
the smallness of $V_{cb}$, the smallness of $m_c/m_t$ compared to
$m_s/m_b$, and the Georgi-Jarlskog factor of three between 
$m_{\mu}$ and $m_s$ at the GUT scale. It should be emphasized that, while
many satisfactory neutrino mixing ideas and also many interesting ideas for 
explaining the pattern of quark and charged lepton masses have been proposed,
very few models exist which not only give a satisfactory
account of neutrino phenomenology but are at the same time highly
predictive.

We have shown that the model defined by the existence of these
five (or six) effective Yukawa terms can be realized in a complete
and specific renormalizable
SUSY $SO(10)$ model that is technically natural. We have presented
the details of such a model, including all the Higgs and quark and lepton
superfields, the abelian flavor symmetries, and the transformation
properties of the fields under these symmetries.  Finally, we
have done a quantitative comparison of the predictions of the model
to experiment. 

In the future this model will be rigorously testable in several ways.
The most important are (1) a relation between the real and imaginary parts of
$V_{ub}$ including a precise test of the angles of the unitarity triangle; 
(2) a prediction for $U_{e2}$, which in the small angle MSW case
gives a sharp relation between the solar and atmospheric angles;
and (3) a definite prediction for $U_{e3}$. 

\vspace*{1in}

The research of SMB was supported in part by Department of Energy Grant 
Number DE FG02 91 ER 40626 A007.  One of us (CHA) thanks the Fermilab 
Theoretical Physics Department for its kind hospitality where much of 
his work was carried out.  Fermilab is operated by Universities Research
Association Inc. under contract with the Department of Energy.
%
%

%
%
\begin{figure}
\centerline{
\epsfxsize=1\hsize
\epsfbox{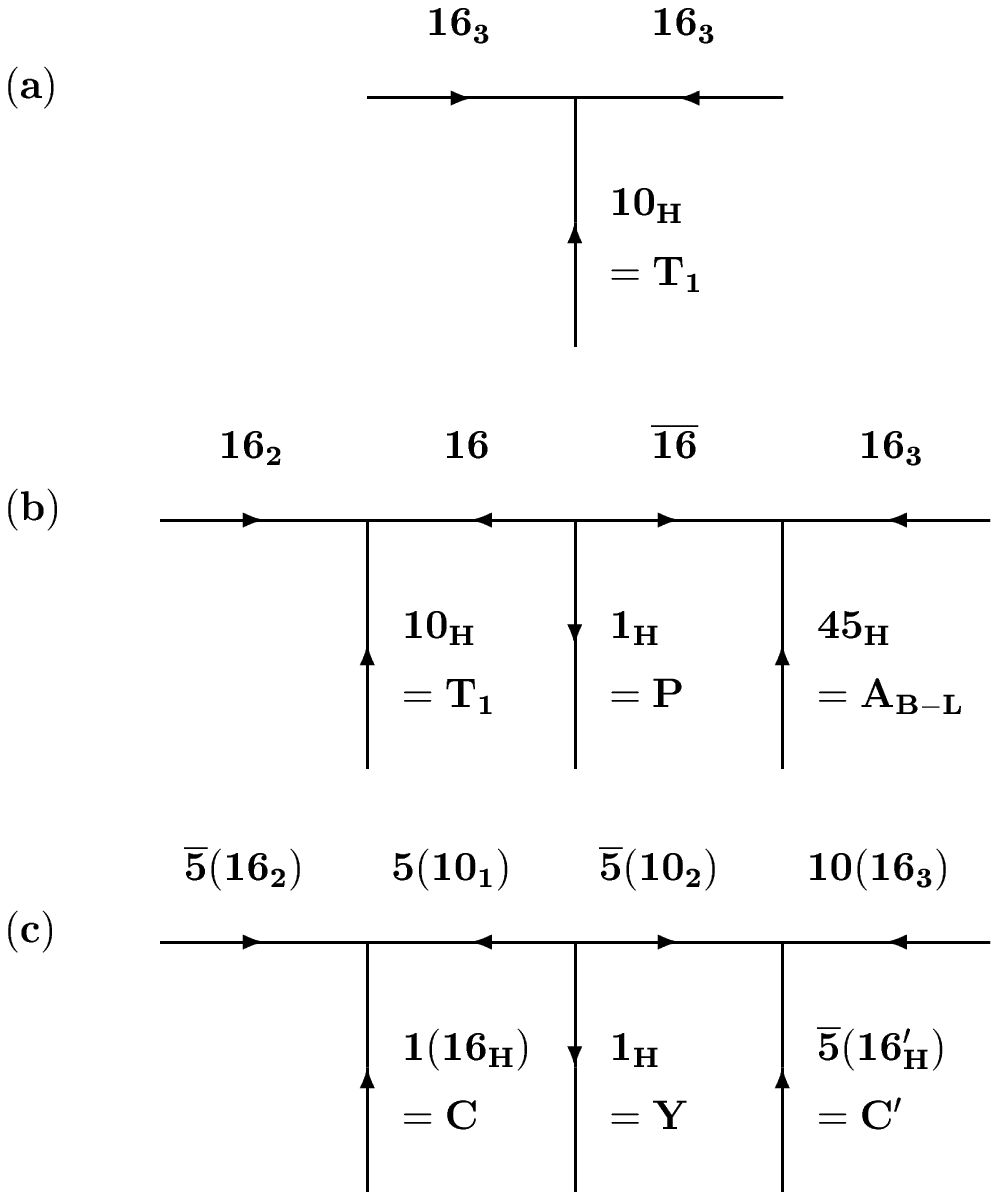}
}
\vspace {0.5in}
\caption{Diagrams that generate the elements in the quark and
	lepton mass matrices shown in Eqs. (10) with the Higgs labeling 
	corresponding to that appearing in Table I of Section VI.
	(a)~The 33 elements denoted ``1". (b) The 23 and 32 elements denoted 
	``$\epsilon$''. Note that because of the appearance of the VEV of
	the adjoint Higgs field ${\bf 45_H} \equiv A$, they are 
	proportional to the $SO(10)$ generator $B-L$.
	(c) The asymmetric elements denoted ``$\sigma$" arise from this
	diagram. That they do not contribute to the up quark masses, and
	contribute asymmetrically to the down quark and charged lepton mass 
	matrices are consequences of the fact that the $SO(10)$ ${\bf 10}$'s,
        i.e., ${\bf 10_1}$ and ${\bf 10_2}$,  contain
	${\bf \overline{5}}$ but not ${\bf 10}$ of $SU(5)$.}
\end{figure}
%
\begin{figure}
\phantom{x}
\centerline{
\epsfxsize=1\hsize
\epsfbox{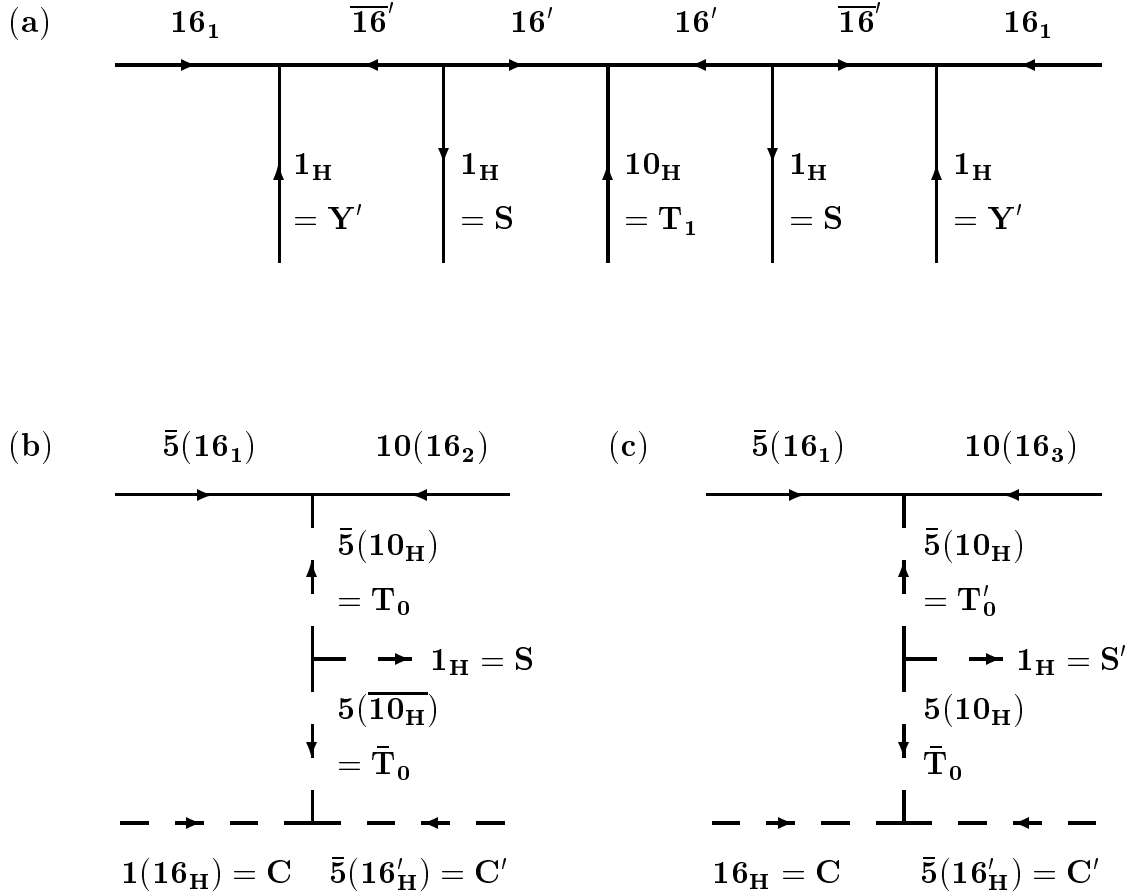}
}
\vspace{0.7in}
\caption{Diagrams that generate the masses of the first family
        of quarks and leptons. See Eq. (14).
	(a) The 11 element called ``$\eta$''.              
        (b) The 12 and 21 elements called ``$\delta$''. 
        Because the ${\bf 10_H}$ couples to the symmetric
        product of ${\bf 16_1}{\bf 16_2}$, $\delta$
        appears symmetrically in the mass matrices.
        (c) The 13 and 31 elements called ``$\delta'$'', which
        also appear symmetrically.}

\end{figure}
\newpage
\begin{figure}
\phantom{x}
\centerline{
\epsfxsize=1\hsize
\epsfbox{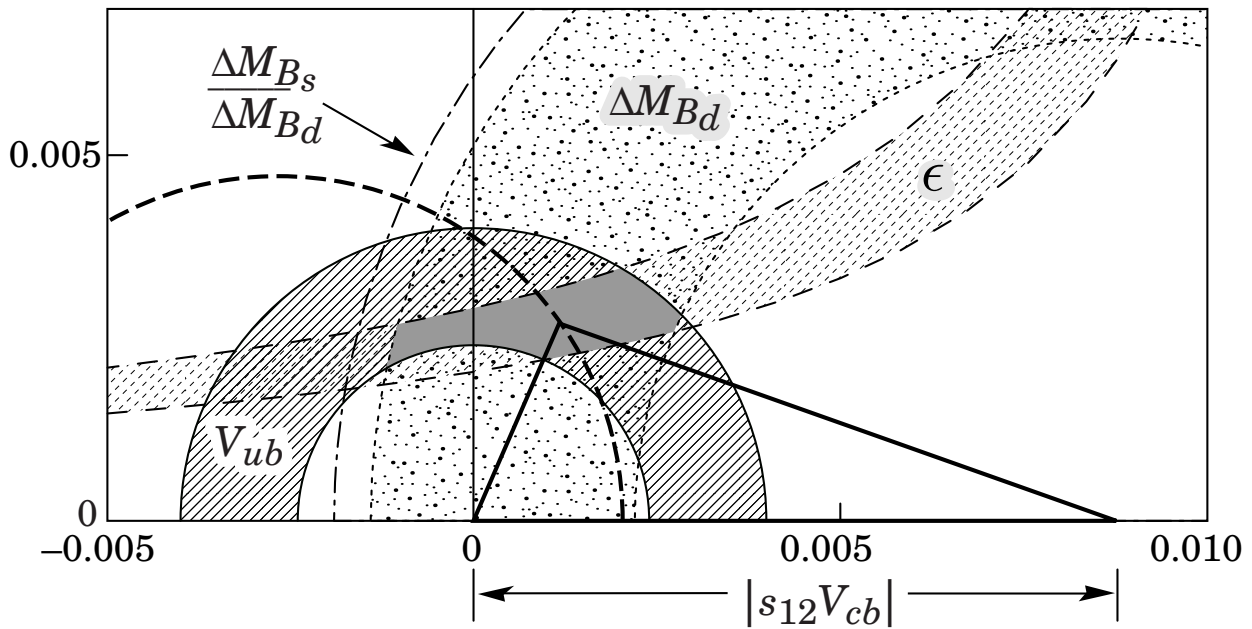}
}
\vspace{0.5in}
\caption{The unitarity triangle for $V_{ud}V^*_{ub} + V_{cd}V^*_{cb}
	+ V_{td}V^*_{tb} = 0$ is displayed along with the experimental 
	constraints
        on $V_{ud}V^*_{ub}$, which is the upper vertex in the triangle.
        The constraints following from
	$|V_{ub}|$, B-mixing and $\epsilon$ extractions from experimental
	data are shown in the lightly shaded regions. The experimentally
	allowed region is indicated by the darkly shaded overlap.
        The model predicts that $V_{ud}V^*_{ub}$ will lie on the dashed
        circle; cf. Eq. (21). The particular point on this circle used to 
	draw the triangle shown is obtained from the numbers given in 
        Section VI; cf. Eq. (62).}
\end{figure}
\newpage
\begin{figure}
\phantom{x}
\centerline{
\epsfxsize=1\hsize
\epsfbox{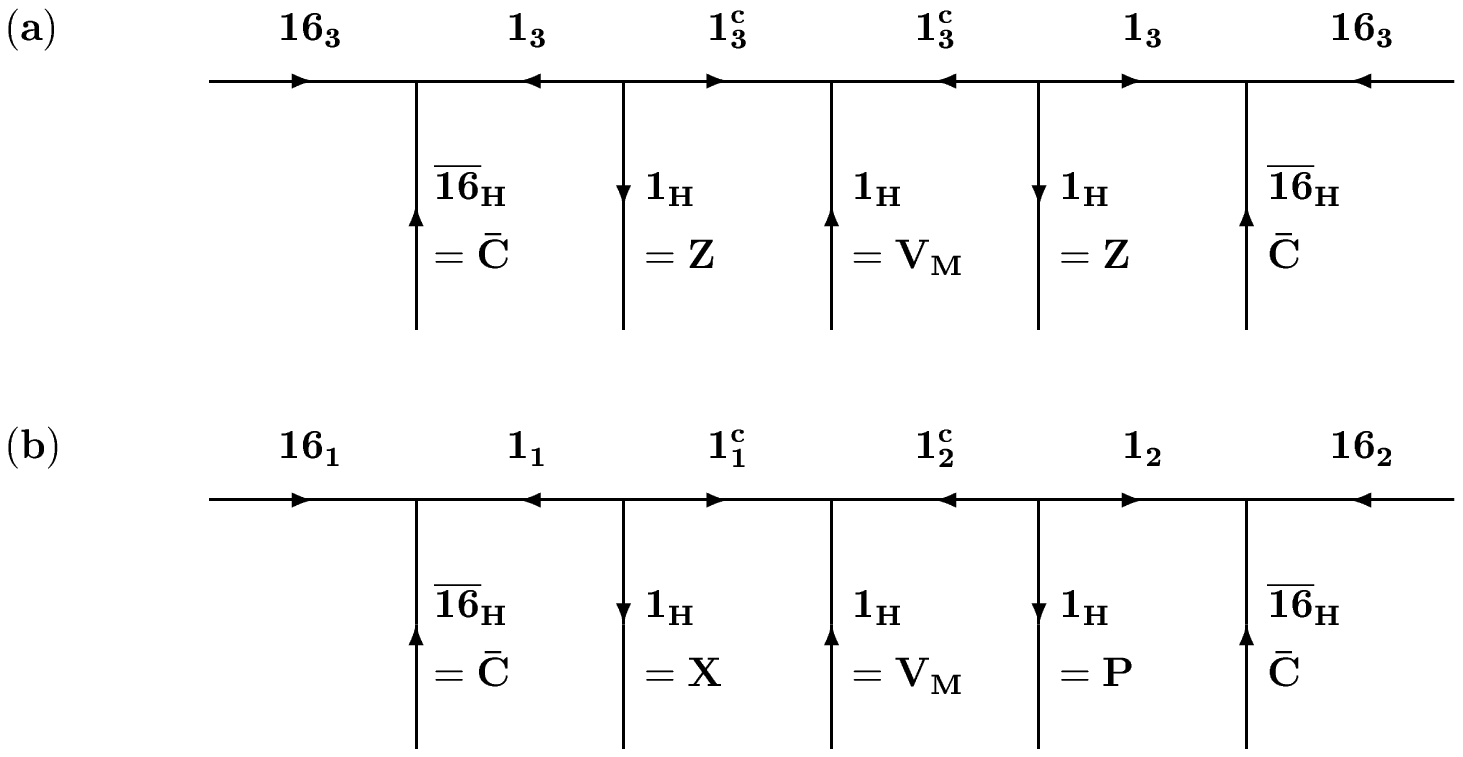}
}
\vspace{-0.5in}
\caption{Diagrams that generate the 33, 12, and 21 elements of the
         Majorana mass matrix $M_R$ of the superheavy right-handed
         neutrinos.}
\end{figure}

\begin{references}
%
\bibitem{ab1} C.H. Albright and S.M. Barr, Phys. Rev. D {\bf 58} 013002
	(1998). 

\bibitem{abb} C.H. Albright, K.S. Babu, and S.M. Barr, Phys. Rev. 
	Lett. {\bf 81}, 1167 (1998); C.H. Albright, K.S. Babu, and 
	S.M. Barr, Nucl. Phys. B (Proc. Suppl.) {\bf 77}, 308 (1999).

\bibitem{ab4} C.H. Albright and S.M. Barr, Phys. Lett. B {\bf 452},
	287 (1999).

\bibitem{ab5} C.H. Albright and S.M. Barr, Phys. Lett. B {\bf 461},
	218 (1999).

\bibitem{b-r} S.M. Barr and S. Raby, Phys. Rev. Lett. {\bf 79}, 
        4748 (1997).

\bibitem{atm}   Y. Fukuda et al., Phys. Rev. Lett. {\bf 81}, 1562 (1998);
                Y. Suzuki, in Proceedings of the WIN-99 Workshop, Cape Town,
                25 - 30 January 1999, to be published.

\bibitem{MSW}   L. Wolfenstein, Phys. Rev. D {\bf 17}, 2369 (1978); S.P. 
                Mikheyev and A. Yu. Smirnov, Yad. Fiz. {\bf 42}, 1441 (1985),
                [Sov. J. Nucl. Phys. {\bf 42}, 913 (1985)].

\bibitem{w-w-z-f} S. Weinberg, Trans. N.Y. Acad. Sci. {\bf 38}, 185 (1977);
	F. Wilczek and A. Zee, Phys. Lett. B {\bf 70}, 418 (1977);
        H. Fritzsch, Phys. Lett. B {\bf 70}, 436 (1977).

\bibitem{Fritzsch} H. Fritzsch, Phys. Lett. {\bf 73B}, 317 (1979).

\bibitem{lop} C.H. Albright, K.S. Babu, and S.M. Barr, Phys. Rev. Lett.
         {\bf 81}, 1167 (1998);
         J. Sato and T. Yanagida,
         Phys. Lett. B {\bf 430}, 127 (1998); N. Irges, S. Lavignac,
         and P. Ramond, Phys. Rev. D {\bf 58}, 035003 (1998).
         See also K.S. Babu and S.M. Barr, Phys. Lett. B {\bf 381}, 202
         (1996).

\bibitem{so10} H. Georgi, {\it Particles and Fields 1974}, ed. C.E. Carlson,
         (AIP, NY, 1975), p.575; H. Fritzsch and P. Minkowski,
         Ann. Phys. {\bf 93}, 193 (1975); R. Barbieri, D.V. Nanopoulos,
         G. Morchio, and F. Strocchi, Phys. Lett. B {\bf 90}, 91 (1980);
         J.A. Harvey, P. Ramond, and D.B. Reiss, Nucl. Phys. B {\bf 199},
         223 (1982).

\bibitem{dim-wil}   S. Dimopoulos and F. Wilczek, report No. NSF-ITP-82-07 
	(1981), in {\it The unity of fundamental interactions}, Proceedings 
        of the 19th Course of the International School of Subnuclear 
        Physics, Erice, Italy, 1981, ed. A. Zichichi (Plenum Press, 
        New York, 1983); K.S. Babu and S.M. Barr, Phys. Rev. D 
        {\bf 48}, 5354 (1993).

\bibitem{I3R} G. Dvali and S. Pokorski, Phys. Lett. B {\bf 379}, 126 (1996);
	S.M. Barr, Phys. Rev. D {\bf 59}, 015004 (1999).

\bibitem{g-j} H. Georgi and C. Jarlskog, Phys. Lett. {\bf B86}, 297
        (1979); A. Kusenko and R. Shrock, Phys. Rev. {\bf D49}, 4962
        (1994).

\bibitem{adjointterm} S.M. Barr, Phys. Rev. Lett. {\bf 64}, 353 (1990);
         K.S. Babu and S.M. Barr, Phys. Rev. Lett. {\bf 75}, 2088 (1995).

\bibitem{bpw} K.S. Babu, J. Pati, and F. Wilczek, Nucl. Phys. B {\bf 566},
         33 (2000).

\bibitem{bando} M. Bando, T. Kugo, and K. Yoshioki, Phys. Rev. Lett.
                {\bf 80}, 3004 (1998). See also K.S. Babu and Q. Shafi,
                Phys. Lett. B {\bf 294}, 235 (1992).

\bibitem{MNS} Z. Maki, M. Nakagawa, and S. Sakata, Prog. Theor. Phys.,
              {\bf 28}, 870 (1962).
 
\bibitem{bimax} H. Fritzsch and Z.Z. Xing, Phys. Lett. B {\bf 372}, 265
                (1996); E. Torrente-Lujan, Phys. Lett. B {\bf 389}, 557 (1996);
		V. Barger, P. Pakvasa, T.J. Weiler, and K. Whisnant, Phys.
                Lett. B {\bf 437}, 107 (1998); A. Baltz, A.S. Goldhaber and 
                M. Goldhaber, Phys. Rev. Lett. {\bf 58}, 5730 (1998);
                H. Georgi and S.L. Glashow, hep-ph/9808293;
                C. Giunti, Phys. Rev. D {\bf 59}, 077301 (1999).

\bibitem{bimaximal} 
                Y. Nomura and T. Yanagida, Phys. Rev. D {\bf 59}, 017303 (1998);
		G. Altarelli and F. Feruglio, Phys. Lett. B {\bf 
                439}, 112 (1998); M. Jezabek and Y. Sumino, Phys. Lett. B 
                {\bf 440}, 327 (1998);  H. Fritzsch and Z. Xing, Phys. Lett. B 
                {\bf 440}, 313 (1998); R. Mohapatra and S. Nussinov, Phys. 
                Lett. B {\bf 441}, 299 (1998); M. Tanimoto, Phys. Rev. D 
                {\bf 59}, 017304 (1999); S.K. Kang and C.S. Kim,
                Phys. Rev. D {\bf 59}, 091302 (1999); C. Jarlskog, M. Matsuda,
                S. Skadhauge, and M. Tanimoto, Phys. Lett. B {\bf 449}, 240
                (1999); Y.-L. Wu, hep-ph 9901245;  E. Ma, hep-ph 9902392;
                A. Ghosal, hep-ph 9905470.

\bibitem{f-n}	C.D. Froggatt and H.B. Nielsen, Nucl. Phys. B {\bf 147},
		277 (1979).

\bibitem{data}	{\it Table of Particle Properties}, Particle Data Group,
		(1998); M. Bargiotti et al., hep-ph/0001293.

\bibitem{CHOOZ}	CHOOZ Collab. (M. Apollonio et al.), Phys. Lett. B {\bf 420},
		397 (1998).

\end{references}
\end{document}